\documentclass[prd,showpacs,twocolumn,nofootinbib]{revtex4}
\usepackage{graphicx}
\usepackage{psfrag}
\usepackage{amsmath}
\usepackage{amsfonts}
\usepackage{amssymb}

\def\be{\begin{equation}}
\def\ee{\end{equation}}
\def\ba{\begin{eqnarray}}
\def\ea{\end{eqnarray}}
\def\bs{\begin{subequations}}
\def\es{\end{subequations}}
\def\half{\frac12}
\def\tphi{\tilde\phi}

\def\vp{\varphi}

\def\TT{{\cal T}}
\def\B{\Box}
\def\a{\alpha}
\def\b{\beta}

\def\s{\sigma}

\def\p{\partial}
\def\ve{\varepsilon}
\def\rme{e}
\def\rmd{d}
\def\rmi{i}
\def\cO{{\cal O}}
\def\Im{{\rm Im}}

\def\nn{\nonumber}
\def\rno{\stackrel{\scriptscriptstyle{\star}}{\scriptscriptstyle{\star}}}
\def\ono{\stackrel{\scriptscriptstyle{\circ}}{\scriptscriptstyle{\circ}}}
\newcommand{\Eq}[1]{(\ref{#1})}

\begin{document}

\title{Tachyon solutions in boundary and open string field theory}
\author{Gianluca Calcagni}
\affiliation{Department of Physics and Astronomy, University of Sussex, Brighton BN1 9QH, United Kingdom}
\affiliation{Institute for Gravitation and the Cosmos, Department of Physics,\\ The Pennsylvania State University,
104 Davey Lab, University Park, PA 16802, USA}
\author{Giuseppe Nardelli}
\affiliation{Dipartimento di Matematica e Fisica, Universit\`a Cattolica, via Musei 41, 25121 Brescia, Italia}
\affiliation{INFN Gruppo Collegato di Trento, Universit\`a di Trento, 38050 Povo, Trento, Italia}
\date{June 1, 2008}

\begin{abstract}
We construct rolling tachyon solutions of open and boundary string field theory (OSFT and BSFT, respectively), in the bosonic and supersymmetric (susy) case. The wildly oscillating solution of susy OSFT is recovered, together with a family of time-dependent BSFT solutions for the bosonic and susy string. These are parametrized by an arbitrary constant $r$ involved in solving the Green equation of the target fields. When $r=0$ we recover previous results in BSFT, whereas for $r$ attaining the value predicted by OSFT it is shown that the bosonic OSFT solution is the derivative of the boundary one; in the supersymmetric case the relation between the two solutions is more complicated.
This technical correspondence sheds some light on the nature of wild oscillations, which appear in both theories whenever $r>0$.
\end{abstract}


\preprint{arXiV:0708.0366 [hep-th] \hfill Phys. Rev D {\bf 78}, 126010 (2008)}

\maketitle


\section{Introduction}

Since the seminal papers by Sen on the rolling tachyon~\cite{se021,se023,se024}, much work has been devoted to the study of time-dependent solutions in string theory. These solutions describe a system of unstable $D$-branes which decay into closed strings as the tachyon field rolls down from the maximum of the potential towards the stable minimum \cite{kle03,dou03}. Besides the boundary state description originally used in \cite{se021}, there are two main approaches to the study of rolling tachyon solutions: boundary string field theory (BSFT) \cite{wit92,wi92b,sh93a,sh93b,sug02,min02,lar02} and open string field theory (OSFT, often called ``cubic'' as the bosonic version has a cubic interaction)~\cite{wi86a,KS1,KS2,fuj03}. While in BSFT rolling tachyons are well established, in OSFT these solutions have been searched for a long time with unsatisfactory results; i.e., even at lowest-level truncation order a smooth solution of the equations of motion interpolating between the two inequivalent vacua was not found, and the supposed equivalence between BSFT and OSFT was in doubt. Two seemingly contrasting results were found: an even bump, nonanalytic at the origin \cite{FGN}, and a solution with wildly increasing oscillations \cite{MZ,CST,FGN2}. Recently, the understanding of OSFT has been improved thanks to the choice of a gauge alternative to the Siegel gauge \cite{sch05}, which allowed one to prove Sen's conjecture analytically \cite{sch05,FK1,ES} (see also \cite{oka06,FK2,FNS}). In particular, the problem of finding bosonic rolling solutions has been reexamined \cite{sch07,KORZ,ell07}, and the existence of an oscillating unbounded solution confirmed.\footnote{For a solution of supersymmetric (Berkovits') SFT, see \cite{erl07,oka1,oka2}. For a bosonic solution in another gauge, see \cite{FKP}.}

The aim of the present paper is threefold. 

\begin{enumerate}
\item[(i)] First, to present OSFT as an example in Minkowski spacetime of a class of nonlocal scalar field theories which can be solved by a method based on the diffusion equation. The wildly oscillating solution of supersymmetric OSFT is naturally recovered and its series and integral representations are given. 
\item[(ii)] Second, to clarify the relation between (and interpretation of) even solutions \cite{FGN}, which are now reinterpreted, and those with wild oscillations. Although they seem not to be in contrast to each other (analytic continuation of even solutions at negative time $t$ to the half plane $t>0$ gives precisely the oscillating behaviour), it is argued that the spiky solution may be unphysical.
\item[(iii)] Last but not least, to clarify why, contrary to what happens in BSFT, the OSFT solution does not describe a condensate in the true vacuum. This is achieved in two steps: (1) a generalization of the BSFT solution to a family depending on a parameter $r$ whose sign (together with the parity of the solution) determines whether a solution in either theory rolls or displays wild oscillations; and (2) a quantitative relation between BSFT and OSFT tachyons: in the bosonic example, the rolling BSFT tachyon is just the antiderivative of the solution of OSFT.
\end{enumerate}

The last property, although it relates approximate solutions of an approximate equation in OSFT with the exact solutions of BSFT, is remarkable and confirms previous findings \cite{ell07}. The rolling solution $\phi(t)$ of $(0,0)$-level bosonic OSFT studied in Ref.~\cite{FGN} is easily written as the series
\be\label{psix}
\psi_-(r,t)=-6\sum_{k=1}^\infty(-1)^k \rme^{-rk^2} k\rme^{kt}\,,
\ee
where $r=r_*\equiv(\ln\lambda_*)/3$, $\lambda_*=3^{9/2}/2^6\approx 2.19$ and $\phi(t)=\lambda_*^{-5/3-\p_t^2/3}\psi(r,t)$.
This equation is an approximate solution of the OSFT equations of motion truncated at the (0,0)-level. Its antiderivative can be written, up to additive and normalization constants, as 
\be\label{ttx}
\vp_-(r,t)=-\sum_{k=1}^\infty (-1)^k\,\rme^{-rk^2} \rme^{kt}.
\ee
We claim and hereafter prove that Eq.~\Eq{ttx} is just a one-parameter family of solutions of BSFT.
Here, the parameter $r$ is arbitrary, not necessarily equal to $r_*$, and reflects an ambiguity in solving the Green equation for this theory. If we set $r=0$, it reproduces the BSFT solutions studied in Refs.~\cite{sug02,min02,lar02}, and if $r$ is left unspecified it provides a generalization of the latter. When $r$ is positive, the BSFT solution has wild oscillations.

The paper is organized as follows. 

The bosonic OSFT case and the diffusion equation method are reviewed in Sec.~\ref{bosmink}. The supersymmetric case is discussed in Sec.~\ref{mink}. The oscillating tachyon solution on a Minkowski target spacetime is described in Sec.~\ref{thesol} and compared with another candidate solution which is an even function nonanalytic at the origin; on physical grounds the latter can be discarded.

In Sec.~\ref{bsftsol} we derive the BSFT bosonic open string disk partition function in the presence of a tachyon profile of the form $T(X)=T_0 \rme^{\rmi p\cdot X}$, where $X^\mu$ are the target scalars and $p_\mu$ is time-like. The calculations are performed by keeping the constant $r$ undetermined. The exact solution is Eq.~(\ref{ttx}). 

The tachyonic solution of supersymmetric BSFT is derived in Sec.~\ref{sbsftsol}. The correspondence between OSFT and BSFT solutions is discussed in Sec.~\ref{bccor}. The last section contains a summary and conclusions.

The appendices are devoted to material which would distract the reader from the main thread. The relation between different representations of the BSFT bosonic solution is shown in Appendix \ref{bsrel} with techniques which can be readily extended to the susy BSFT and OSFT solutions. The rolling solutions in BSFT have a close relationship to the one obtained through boundary states~\cite{se021,se022} (see also \cite{LS,lee06}). In Appendix \ref{bounda} we construct the solution \Eq{ttx} in this framework. There, the presence of the arbitrary parameter $r$ is justified by the order ambiguity in the regularization of quantum correlators. It is always possible to define an $r$ ordering which tends to the usual normal ordering when $r \to 0$.


\section{Bosonic OSFT} \label{bosmink}


\subsection{General setup}\label{bossetup}

The bosonic OSFT action is of Chern--Simons type \cite{wi86a},
\be\label{SFT}
S=-\frac{1}{g_o^2}\int \left(\frac{1}{2\alpha'} \Phi* Q_B\Phi+\frac13\Phi*\Phi *\Phi\right),
\ee
where $g_o$ is the open string coupling constant (with $[g_o^2]=E^{6-D}$ in $D=26$ dimensions), $\int$ is the path integral over matter and ghost fields, $Q_B$ is the BRST operator, * is a noncommutative product, and the string field $\Phi$ is a linear superposition of states whose coefficients correspond to the particle fields of the string spectrum.

At the lowest truncation level \cite{lump2}, all particle fields in $\Phi$ are neglected except the tachyonic one, labeled $\phi(x)$ and depending on the center-of-mass coordinate $x$ of the string. The Fock-space expansion of the string field is truncated so that $\Phi \cong |\Phi\rangle=\phi(x)|\!\downarrow\rangle$, where the first step indicates the state-vertex operator isomorphism and $|\!\downarrow\rangle$ is the ghost vacuum with ghost number $-1/2$.
At level $(0,0)$ the action becomes, in $D=26$ dimensions and with metric signature $({-}{+}{\dots}{+})$ \cite{KS1,KS2},
\ba
\bar{S} &=&\frac{1}{g_o^2}\int \rmd^D x \left[\frac{1}{2\alpha'}\phi(\alpha'\p_\mu\p^\mu+1)\phi\right.\nonumber\\
&&\qquad\left.-\frac{\lambda_*}{3}\left(\lambda_*^{\alpha'\p_\mu\p^\mu/3}\phi\right)^3-\Lambda\right],\label{tactmin}
\ea
where $\lambda_*=3^{9/2}/2^6$ , $\alpha'$ is the Regge slope, and Greek indices run from 0 to $D-1$ and are raised and lowered via the Minkowski metric $\eta_{\mu\nu}$. The tachyon field is a real scalar with dimension $[\phi]=E^2$. The constant $\Lambda$ does not contribute to the scalar equation of motion but it does determine the energy level of the field. In particular, it corresponds to the $D$-brane tension which sets the height of the tachyon potential at the (closed-string vacuum) minimum to zero. This happens when $\Lambda=(6\lambda_*^2)^{-1}$, which is around $68\%$ of the brane tension; this value is lifted up when taking into account higher-level fields in the truncation scheme.

We define the operator
\be\label{lb}
\lambda_*^{\B/3}= \rme^{r_*\B} \equiv\sum_{\ell=0}^{+\infty}\frac{(\ln\lambda_*)^\ell}{3^\ell \ell!} \B^\ell=\sum_{\ell=0}^{+\infty}c_\ell \B^\ell\,,
\ee
where $\Box \equiv -\p_t^2$ and
\be
r_*\equiv \frac{\ln\lambda_*}{3}=c_1=\ln 3^{3/2}-\ln 4\approx 0.2616.
\ee
Defining the ``dressed'' scalar field
\be\label{dres}
\tphi\equiv \lambda_*^{\B/3}\phi=\rme^{r_*\B}\phi,
\ee
the total action is
\be\label{tact}
S=\int \rmd^D x \left[\frac12\,\phi(\B-m^2)\phi-U(\tphi)-\Lambda\right],
\ee
where $m^2$ is the squared mass of the field (negative for the tachyon) and we have absorbed the open string coupling into $\phi$, so that the latter has dimension $[\phi]=E^{(D-2)/2}$.

The equation of motion for the SFT tachyon is (see \cite{yan02,cutac} for the detailed derivation of the dynamical equations)
\be\label{teom}
\B\phi=m^2\phi+U'\,,
\ee
where
\be\label{mono}
U'=\rme^{r_*\B}\tilde U'\equiv\rme^{r_*\B}\frac{\p U}{\p\tphi},\\
\ee
is constructed from a nonlocal potential term $U(\tphi)$ which does not contain derivatives of $\tphi$. One can also recast Eq.~\Eq{teom} in terms of $\tphi$,
\be\label{teom2}
(\B-m^2) \rme^{-2r_*\B}\tphi=\tilde U'.
\ee
When the nonlocal term is a monomial, the total tachyonic potential is
\be\label{pot}
\tilde V(\phi,\tphi) \equiv \frac{1}{2}m^2\phi^2+\frac{\sigma}{n}\tphi^n+\Lambda,
\ee
where $\sigma$ is a coupling constant and we have isolated the quadratic local mass term, with $m^2$ being a dimensionless number, and $\Lambda$ is the (possibly  vanishing) cosmological constant, which sets the energy level
\be\label{energy}
E=\frac{\dot{\phi}^2}{2}(1-\cO_2)+\tilde V -\cO_1,
\ee
where
\ba
\cO_1 &=& \int_0^{r_*} \rmd s\, (\rme^{s\B}\tilde U')(\B \rme^{-s\B}\tphi),\nonumber\\
\cO_2 &=& \frac{2}{\dot{\phi}^2}\int_0^{r_*} \rmd s\, \p_t(\rme^{s\B}\tilde U')\p_t(\rme^{-s\B}\tphi)\,.
\ea
In the local case ($r_*=0$, $\lambda_*=1$), $\cO_i=0$. The tachyon of the bosonic string has 
\be
U(\tphi) = \frac{\lambda_*}{3}\tphi^3\,,\qquad m^2 = -1\label{cupot}\,.
\ee


\subsection{Truncated power-series solution}\label{ps}

As $r=r_*$ is a small number, one can try to find a homogeneous solution as a power series in $r$ (subscript * ignored from now on). The leading term $r=0$ is the solution of the local system, which is
\be\label{bos20}
\phi(0,t) \equiv \phi_{\rm loc}(t)= \frac{3}{2\cosh^2 t/2}=6\int_0^{+\infty}\rmd\s\,\frac{\s \cos (\s t)}{\sinh (\pi\s)}\,,
\ee
where we wrote a useful integral representation. Applying the nonlocal operator, one gets
\ba
\psi(r,t) &=& 6\rme^{r \B}     \int_0^{+\infty}\rmd\s\,\frac{\s \cos (\s t)}{\sinh (\pi\s)}\nonumber\\
&=& 6\int_0^{+\infty}\rmd\s\,\rme^{r\s^2}\ \frac{\s \cos (\s t)}{\sinh (\pi\s)}\,.\label{bos2}
\ea
Expanding the exponential as $\rme^{r\s^2}\approx\sum_{n=0}^{n_{\rm max}} (r\s^2)^n/n!$, $r<0$, Eq.~\Eq{bos2} would display growing oscillations\footnote{These are not to be confounded with the oscillations at $t>0$ of the solution below.} near the origin and diverge at $t=0$. This is a spurious effect of the truncation, and the full expression \Eq{bos2} must be used instead. Although this example is valid only for negative $r$, the same problem reappears in the physical case $r>0$, where there are no oscillations but the function blows up at the origin.


\subsection{Diffusion equation method}\label{GF}

We derive a solution of SFT following the method outlined in \cite{FGN,ctac2}.
The same features encountered in the bosonic case \cite{FGN} will emerge in the supersymmetric string, i.e., solutions with either a spike (a point where the left and right derivatives are finite but different\footnote{The spike was not recognized in \cite{FGN}, whose discussion on the point $t=0$ is now superseded.}) or wild oscillatory behaviour. Since the same strategy can be adopted also in other examples on curved backgrounds \cite{ctac2,CN}, we shall discuss the method in detail.
\begin{enumerate}
\item[(1)] Interpret $r_*$ as a fixed value of an auxiliary evolution variable $r$, so that the scalar field $\phi=\phi(r,t)$ is thought to live in $1+1$ dimensions (there is no role of the spatial directions in this discussion). Find a solution of the corresponding \emph{local} system ($r=r_*=0$ everywhere). This is the initial condition for a system that evolves in $r$.
\item[(2)] Solve the eigenvalue equation of the d'Alembertian operator, $\B G_k(t)=k^2 G_k(t)$.
\item[(3)] Write the local solution ($r=0$) as a linear combination of the eigenfunctions of the d'Alembertian operator
 \be\label{step3}
 \phi(0,t)=\sum_k c_k G_k(t)\,.
 \ee
If the spectrum of $\B$ is continuous, the above series is replaced by an integral in $k$.
\item[(4)] Look for nonlocal solutions $\phi(r,t)$ of the type $\rme^{r (\b +\B/\a)} \phi(0,t)$, for some (unknown) parameters $\a$ and $\beta$. Notice that the action of nonlocal operators of the type $\rme^{(r/\a)\B}$ on the local solution $\phi(0,t)$ now simply corresponds to the replacement $c_k\to \rme^{r k^2/\a} c_k$ in the sum \Eq{step3}. Thus one looks for solutions of the type
\ba
\ \ \phi(r,t)&=&\rme^{r (\b +\B/\a )} \phi(0,t)=\rme^{r \b}\sum_k \rme^{r k^2/\a } c_kG_k(t)\,.\nonumber\\
\label{step4}
\ea
\item[(5)] The coefficients $\alpha$ and $\beta$ such that Eq.~\Eq{step4} is a solution (exact or approximate)  of Eq.~\Eq{aref1} can be chosen either by equating the ``modes'' $G_k(t)$ in the two sides of the equation of motion or by variational techniques.
\end{enumerate}
The great advantage of this procedure is that it makes the equation of motion \emph{local} in the time variable $t$; the $(1+1)$ system solved by some $\phi(r,t)$ will be referred to as \emph{localized}. By construction, $\psi(r,t)\equiv\rme^{-\b r} \phi(r,t)$ satisfies the homogeneous diffusion equation
\be \label{step5}
\a\p_r \psi(r,t)= \B\psi(r,t)\,.
\ee
As a consequence
\be\label{tra}
\rme^{q\B}\psi(r,t)= \rme^{\a q\,\p_r} \psi(r,t)=\psi(r+\a q,t)\,,
\ee
and the effect of the nonlocal operator $\rme^{q\B}$ is a shift of the auxiliary variable $r$. In our case, $q$ must be a multiple of $r$; since $r$ and $\p_r$ do not commute, we need an ordering prescription for the exponential. We adopt the one compatible with the diffusion equation \Eq{step5}, setting all the derivatives $\p_r$ to the right of the powers of $r$. In fact, 
\ba
\rme^{r\B}\psi(r,t)&=&\sum_{k=0}^\infty \frac{r^k }{k!}\B^k \psi(r,t)=\sum_{k=0}^\infty \frac{(\a r)^k }{k!}\p_r^k \psi(r,t)\nonumber\\
&=&\psi((1+\a)r, t)\,.\label{tra2}
\ea
Then one can check whether the found solution fulfils the equation of motion globally (that is, at all times) or locally (i.e., in any specified time interval). This check is not possible if the nonlocal operator is expanded as a truncated power series, as any such analysis would be necessarily limited only to solutions of the form $\tphi=(1+c_1\B+\dots+c_{\ell_{\rm max}}\B^{\ell_{\rm max}})\phi$. In other words, one can only increase the truncation order $\ell_{\rm max}$ and see numerically whether the solution is convergent and, in this case, sensibly postulate that the fully resummed solution enjoys the same properties of the truncated one. However, the sum is unknown and a formal proof of global or local convergence is not possible. On the other hand, there is no issue of convergence for localized systems.


\subsection{Rolling solutions of bosonic cubic SFT}\label{bossol}

An even solution of bosonic OSFT was found in \cite{FGN} and here we recall the main equations; they can be easily derived via the methods below. The series representation for $r>0$ is Eq.~\Eq{psix} for $t<0$ and $\psi_+(r,t)=\psi(r,-t)$ for $t >0$, the integral representation for $r<0$ is Eq.~\Eq{bos2}, while for $r>0$ one has\footnote{Equation \Eq{regux} corresponds to Eq.~(2.26) in \cite{FGN}, integrated twice by parts, with $\ln\lambda\to(\ln\lambda)/3$ in order to match our conventions. Compare Eq.~(2.12) with Eq.~\Eq{step4} below.}
\be
\psi(r,t)=-6\int_0^\infty \rmd\s\ \p_\s K(\s,r)\,\frac{\sin \s}{\rme^\ve \cosh t + \cos\s}\,,\label{regux}
\ee
where 
\be
\label{ker}
K(\s,r)=\frac{\rme^{-\frac{\s^2}{4r}}}{2\sqrt{\pi r}}\,.
\ee
The $\rme^\ve$ term redefines $\psi$ and all its derivatives at the origin and it can be removed after integration over $\s$ is performed. Note that the integral Eq.~\Eq{regux} with $\ve=0$ (``strong limit'') is ill-defined in $t=0$, as the integral picks up the poles of the integrand  at $\sigma= \pi(2k+1)$. Hence one would conclude that $\psi(0)=\infty$, while the spike does have a well-defined finite value. This suggests that the ``weak limit'' (first integrate, then set $\ve\to 0$) is the only one in which the integral representation of $\psi$ makes any sense.\footnote{A simple example of strong and weak limits is provided by the Fourier transform of the retarded distribution. As a weak limit, it is just the definition of the Heaviside (step) function, $\Theta (p)=
\lim_{\ve\to 0}(2 \rmi\pi)^{-1}\int_{-\infty}^{+\infty}\rmd x\,\rme^{\rmi px}/(x-\rmi\ve)$. However, setting $\int\rmd x\cos(px)/x =0$ by symmetry, the integral with $\ve=0$ also exists (strong limit), and the result would be $(1/2){\rm sgn}(p)$, which is the Fourier transform of the principal value distribution. The difference of the two results is a constant, which is a contact ($\delta$) term in the integrand. The same phenomenon happens also in our case
and the choice of limiting procedure, made at the level of the solution $\psi$ of the equation of motion, is dictated by the physics of the problem.} The discussion of these formul\ae\ will be amended with respect to the material presented in \cite{FGN}. Since it runs along the same lines as for the susy case, we postpone it to the next section.

The wildly oscillating solution is simply the analytic continuation of Eq.~\Eq{regux} from $t<0$ to positive times, Eq.~\Eq{psix} (we will soon expand this statement).


\section{Supersymmetric OSFT} \label{mink}


\subsection{General setup}\label{setup}

Contrary to the cubic string, there are several proposals for superstring field theory, the first being
Witten's \cite{wi86b,con1,con2,con3,con4,DR}. The action was later modified by \cite{AMZ1,AMZ2,PTY} as
\be\label{SSFT}
S=-\frac1{g_o^2}\int Y_{-2}\left(\frac1{2\alpha'}\Phi* Q_B\Phi+\frac13\Phi*\Phi*\Phi\right),
\ee
where $Y_{-2}$ is a double-step inverse picture-changing operator and $\Phi$ now includes superfields in the 0-picture.
The operator $Y_{-2}$ can be either chiral and local 
\cite{AMZ1,AMZ2} or nonchiral and bilocal 
\cite{PTY}  (see the literature and the review 
\cite{ohm01} for full details). These two theories  predict the same tree-level on-shell amplitudes but different off-shell sectors. 

From now one we concentrate on the nonchiral version \cite{PTY,AKBM}. At level $(1/2,1)$, which is the lowest for the susy tachyon effective action, the tachyon potential is \cite{AJK}
\be
U(\tphi) = \frac{\rme^{4r_*}}{36} \left(\rme^{r_*\B}\tphi^2\right)^2\,, \qquad \label{4pot}
m^2 = -1/2\,.
\ee
Equation \Eq{4pot} contains derivatives of $\tphi$ and Eq.~\Eq{mono} does not apply. Rather, the susy equation of motion is
\be\label{eox}
(\B-m^2) \phi=U'=\s \rme^{r_*\B}(\tphi\,\rme^{2r_*\B}\tphi^2)\,,
\ee
but at first we will use the approximation \cite{AJK}
\be\label{apex}
\rme^{2r_*\B}\tphi^2\approx \tphi^2\,,
\ee
in order to have a qualitative idea about the behaviour of the supersymmetric string. Below we verify that 
the nonlocal solution of the approximated system Eq.~\Eq{apex} is not a solution of Eq.~\Eq{eox}, but it 
will be straightforward to find the latter.


\subsection{Rolling solutions of supersymmetric OSFT}\label{thesol}

Rescaling $t\to \sqrt{2}t$ and $\phi\to 3 \phi$ in Eqs.~\Eq{eox} and \Eq{apex}, the (approximate) susy equation of motion for a purely homogeneous field configuration reads
\be\label{aref}
(1-\p^2_t)\phi=2 \rme^{4r_*}\rme^{-\frac{r_*}{2}\p^2_t}\bigl(\rme^{-\frac{r_*}{2}\p^2_t} \phi\bigr)^3\,.
\ee
Performing the field redefinition $\bar\phi=\rme^{\frac{r_*}{2}\p^2_t}\phi$, and neglecting the bar over $\phi$ to keep notation light, Eq.~\Eq{aref} becomes
\be\label{aref1}
(1-\p^2_t)\phi=2 \rme^{4r_*}\bigl(\rme^{-r_*\p^2_t} \phi\bigr)^3\,,
\ee  
so that
\be
\sigma=2 \lambda_*^{4/3}=2\rme^{4r_*}\,,\qquad m^2=-1\,,
\ee
in Eq.~\Eq{pot}.

Let us follow the recipe of Sec.~\ref{GF} step by step. A solution with $r=0$ satisfying the boundary condition $\phi(r=0, t=-\infty)=0$ is $\phi(0,t)=\pm\,{\rm sech}t$, where the $\pm$ sign reflects the degeneracy of the potential under the exchange $\phi\to -\phi$. From now on and without loss of generality, we shall consider the positive sign, corresponding to the rolling of the tachyon to the right side of the potential. The eigenfunctions of $-\p^2_t$ are obviously $\rme^{\rmi k t}$. The local solution $\phi(0,t)$ can be easily expanded on the basis of these eigenfunctions. However, the explicit expansion depends on the sign of the eigenvalues $k^2$. Accordingly, Eq.~\Eq{step3} splits into two distinct cases. If $k^2>0$, the sum in Eq.~\Eq{step3} becomes an integral and it provides the Fourier expansion of ${\rm sech}t$:
\be
\label{loc1}
\phi(0,t)={\rm sech}t = \frac{1}{2}\int_{-\infty}^{+\infty} \rmd \s\ \frac{\cos (\s t)}{\cosh (\pi \s/2)}\,.
\ee
If $k^2<0$, Eq.~\Eq{step3} gives the expansion of ${\rm sech}t$ as a geometric series. Convergence of these series imposes two different representations depending on the sign of $t$,
\ba
\  \phi_+(0,t)=\frac{2}{\rme^t+\rme^{-t}}=2 \sum_{k=0}^\infty(-1)^k \rme^{-(2 k +1)t },\;\;
\;\; t >0\,,\cr
\phi_-(0,t)=\frac{2}{\rme^t+\rme^{-t}}=2 \sum_{k=0}^\infty(-1)^k \rme^{(2 k +1)t },\;\;\;\;
\;\; t <0\,,\nonumber\\
\label{loc2}
\ea
where $k$ has been redefined to be real. Notice that, strictly speaking, none of the sums in Eq.~\Eq{loc2} is defined at $t=0$; the value $\phi(0,0)=1$ are defined by analytic continuation of any of the sums. 

Next, applying $\rme^{r (\b-\p^2_t/\a )}$ to Eqs.~\Eq{loc1} and \Eq{loc2} we get
\be
\phi(r,t) 
= \frac{\rme^{\b r}}{2} \int_{-\infty}^{+\infty} \rmd \s\, \rme^{r \s^2/\a}\frac{\cos (\s t)}{\cosh (\pi \s/2)}\,,\label{nloc1}
\ee
and
\ba
\phi_+(r,t)&=&2\rme^{r \b}\sum_{k=0}^\infty(-1)^k \rme^{-r (2k+1)^2/\a } \rme^{-(2 k +1)t},\;\;
\; t >0\,,\nonumber\\
\phi_-(r,t)&=&2\rme^{r \b}\sum_{k=0}^\infty(-1)^k \rme^{-r (2k+1)^2/\a } \rme^{(2 k +1)t},\;\;\;\;
 t <0\,.\nonumber\\
\label{nloc2}
\ea
The Gaussian factors must have the appropriate signs in order for Eqs.~\Eq{nloc1} and \Eq{nloc2} to be well-defined. Choosing $\a>0$, Eq.~\Eq{nloc1} is defined for $r<0$ and Eq.~\Eq{nloc2} for $r>0$; this sign choice is justified \emph{a posteriori} noting that the equation of motion is not solved even approximately when $\a<0$. For $r<0$, $\phi(r,t)\in C^\infty$ [Eq.~\Eq{nloc1}], whereas if $r>0$, $\phi(r,t)$ presents a spike at the point $t=0$, Eq.~\Eq{nloc2} (for any other $t$, it is $C^\infty$). 
The two cases behave differently because $\phi(r,t)$ satisfies the diffusion equation with negative diffusion coefficient. Since the ``initial condition'' in $r$ has been given for $r=0$, the diffusion flow is for negative values of $r$. In Eq.~\Eq{nloc2}, on the contrary, the evolution in $r$ is opposite to the natural flow and a nonanalytic point is expected on general grounds. The physical case is obtained for $r=r_*\approx 0.26$, so we shall have to consider Eq.~\Eq{nloc2}.

The final step is to fix the values of $\alpha$ and $\beta$ such that the equation of motion \Eq{aref1} is approximately satisfied. One can either minimize the $L_2$ norm of the equation of motion with respect to the parameters $\alpha$ and $\beta$ or, more simply, impose that the first coefficients of the modes $\rme^{(2 k +1)t}$  in the expansion of the left- and right-hand sides (LHS and RHS, respectively) of Eq.~\Eq{aref1} coincide.\footnote{This truncation at finite $k$ is of a very different nature with respect to the truncation of the series operator $\rme^\B$: in the former case, this operator is fully resummed.}
In either case, the answer is
\be\label{ab}
\a=1\,,\qquad \b=-7/2
\ee
(for the first two coefficients of the series; by including the third, $\alpha$ and $\beta$ change less than 10\%). In order to avoid confusion in the derivative with respect to $r$ in the diffusion equation, one absorbs the factor in $\b$ redefining
\be\label{psi}
\psi(r,t)=\rme^{7r/2}\phi(r,t)\,.
\ee
Notice that the local version of the two functions coincides, $\psi(0,t)=\phi(0,t)$. Then, taking into account Eq.~\Eq{tra}, the equation of motion becomes local in the variable $t$,
\be\label{aref2}
(1-\p^2_t)\psi(r,t)=2 \rme^{-3r}[\psi(2r,t)]^3\,,
\ee  
and its approximate (although very accurate) solution is, for $r>0$,
\ba
\  \psi_+(r,t)=2\sum_{k=0}^\infty(-1)^k \rme^{-r(2k+1)^2} \rme^{-(2 k +1)t},\;\;\;\; t >0\,,\cr
\psi_-(r,t)=2\sum_{k=0}^\infty(-1)^k \rme^{-r(2k+1)^2} \rme^{(2 k +1)t},\;\;\;\;\;\; t <0\,.
\label{nloc3}
\ea
Besides the equation of motion \Eq{aref2}, $\psi(r,t)$ satisfies the diffusion equation \Eq{step5} with $\alpha=1$. Unfortunately, at the origin $t=0$ the second derivative of Eq.~\Eq{nloc3} develops a $\delta$-function, so the solution breaks down at that point. One can show that there is a way to circumvent this problem by defining an even integral representation of $\psi$, analog to Eq.~\Eq{regux}, which coincides with Eq.~\Eq{nloc3} almost everywhere, is not singular in $t=0$ upon derivation, and is a very accurate global solution of the equation of motion:
\ba
\psi(r,t) &=&\lim_{\ve\to 0}\psi_\ve(r,t)\nonumber\\
&=&\lim_{\ve\to 0}\int_{-\infty}^{+\infty}\rmd\s\,K(\s,r)\,\frac{\cos\s}
{\rme^\ve\cosh t + \sin \s}\,,\label{ana}
\ea
where the limit is performed after integration. To understand to what extent Eq.~\Eq{nloc3} is a solution of Eq.~\Eq{aref2}, we can evaluate the $L_2$ norm of Eq.~\Eq{aref2} written in the form $({\rm LHS}-{\rm RHS})$ and compare it with a typical scale in the problem, that is the $L_2$ norm of $\psi$ or $({\rm LHS}+{\rm RHS})$ (both are of the same order). The result evaluated at $r=r_*$ is $\Delta\equiv\int_{-\infty}^{+\infty}\rmd t({\rm LHS}-{\rm RHS})^2/\int_{-\infty}^{+\infty}\rmd t({\rm LHS}+{\rm RHS})^2\sim 10^{-8}$. It is also easy to check whether Eq.~\Eq{nloc3} is a solution also in the exact case, Eq.~\Eq{4pot}. The equation of motion \Eq{aref2} becomes $(1-\p^2_t)\psi(r,t)=2 \rme^{(4+2\b)r}\psi[(1+\a)r,t]\,\rme^{-r\p^2_t}\psi^2[(1+\a)r,t]$. With the same values of Eq.~\Eq{ab}, Eq.~\Eq{nloc3} is not a solution at any time. This shows that Eq.~\Eq{apex} is not, for this global solution, a good approximation.\footnote{It is possible that the approximation Eq.~\Eq{apex} is valid for other (e.g. kink-type) solutions asymptotically \cite{AJK}.} The equation of motion can be expanded in powers of time and written as $\sum_n a_n \rme^{-nt}=0$. Imposing $a_n=0$ for the first $n$'s, the (approximated) solution 
is given by Eq.~\Eq{nloc2} with $\a\approx 0.67330\approx \frac23$, $\b\approx -2.95564 \approx -3$,
which gives $\Delta\sim 10^{-13}$. Therefore this global solution can be 
considered as exact for all purposes.

\begin{figure}\begin{center}
\includegraphics[width=8.6cm]{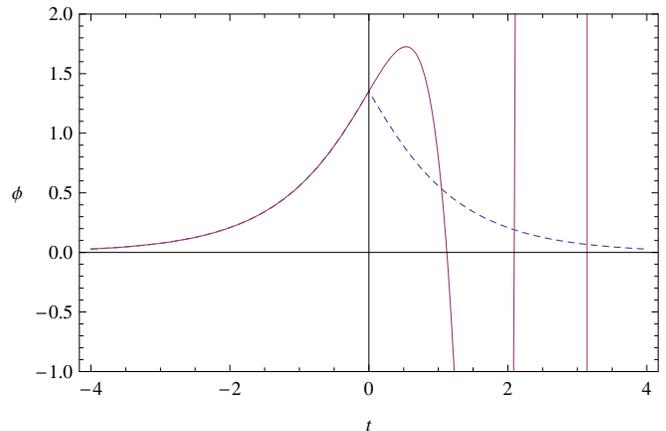}
\caption{\label{fig1} The approximated solutions of the nonlocal approximate supersymmetric system. Solid curve: the wild oscillatory solution Eq.~\Eq{nloc3x}. Dashed curve: Eq.~\Eq{nloc3}, which coincides with Eq.~\Eq{ana}. The series are truncated at $k\sim 10^2$. The spike is at $\psi(r_*,0)\approx 1.3526$. The figure is unchanged for the solution of the system with nontrivial nonlocal potential, except for the height of the spike (lowered down to $1.2956$).}
\end{center}\end{figure}

However, we will eventually reject it and therefore omit a lengthy technical discussion of Eq.~\Eq{ana}. In fact, the splitting of $\psi$ in two series is due to the convergence condition of the local series Eq.~\Eq{nloc2}. When the Gaussian factor is introduced, one is entitled to select only one branch, as the Gaussian factor $\rme^{-r(2k+1)^2}$ has the effect of enlarging the convergence abscissa to the whole real axis. The only constraint is given by the boundary condition at $t=-\infty$ (rolling from the local maximum), so another solution is given by $\psi_-$ with domain extended to positive values of $t$:
\be
\psi_-(r,t)=2\sum_{k=0}^\infty(-1)^k \rme^{-r(2k+1)^2} \rme^{(2 k +1)t}.\label{nloc3x}
\ee
Thus, Eq.~\Eq{ana} can be regarded as a solution with the particular future boundary condition $\psi(r,+\infty)=\psi_+(r,+\infty)=0$. Figure \ref{fig1} shows the two alternative solutions.
Equation \Eq{nloc3x} admits an integral representation, as $k^2<0$ (exponential rather than pure-phase eigenvalues) naturally leads to a Laplace transform which has no definite parity. Starting from the second line of Eq.~\Eq{loc2},
\ba
\psi_-(0,t)&=& \int_0^\infty ds \sum_{k=-\infty}^\infty \rme^{s t -i \frac{\pi}{2}(s-1)} \delta[s-(2 k + 1)]\cr
&=&2 \pi \int_0^\infty ds\  \rme^{s t -i \frac{\pi}{2}(s-1)} \delta\left[\sqrt{2(1+\cos\pi s})\right],\nonumber\\\label{nloc3y}
\ea
where in the first line we have extended the sum to negative $k$'s as these are are ignored by the integral domain.
In the second line we have used the definition of periodic $\delta$ function. This can be written as the limit
\ba
2 \pi \delta \left[\sqrt{2 \left(1+ \cos\pi s \right)}\right]&=&
\lim_{\ve \to 0}\frac{\ve}{1+\ve^2/2+ \cos\pi s}\cr
&=&\lim_{\ve\to0}\frac{\sinh \ve}{\cosh\ve+\cos\pi s}\,.\label{bbbb}
\ea
Substituting Eq.~\Eq{bbbb} in \Eq{nloc3y}, one obtains
\be
\psi_-(0,t)=\lim_{\ve\to0}  \int_0^\infty ds\  \rme^{s t} \frac{\sinh \ve\,\sin(\pi s/2)}{\cosh\ve+\cos\pi s}\,,
\ee
which is well-defined if $t$ is negative. However, applying the nonlocal operator convergence is expanded on the whole real axis:
\be\label{irx}
\psi_-(r,t)=\lim_{\ve\to0}  \int_0^\infty ds\  \rme^{s t-r s^2} \frac{\sinh \ve\,\sin(\pi s/2)}{\cosh\ve+\cos\pi s}\,.
\ee
This function does coincide with Eq.~\Eq{psix}, as one can see also by numerical plots. The same result is achieved directly by taking the Laplace antitransform of $\psi(0,t)$, which is the integral (closed to the left if $t<0$) in $\rmd t$ of $e^{-st}{\rm sech}t$ on the line $\gamma\pm i \infty$, $\gamma>0$. The infinitely many poles of the integrand at $t_k= i\pi(2 k+1)/2$ have residue
\be
R_k=
{\rm Res}\left(\frac{e^{-s t}}{\cosh t} \right)_{t_k}=(-1)^k e^{-i {\pi s/2 }}e^{{-i \pi s  k}}\,.
\ee
This is a sum of phases which is regularized in a standard way by inserting a convergence factor $e^{\pm \ve}$ (depending of the sign of $k$): 
\ba
\sum_{k=-\infty}^\infty R_k &=& \rme^{-i {\pi s/2 }}\sum_{k=0}^\infty (-1)^k e^{-\ve k - i \pi s k}\nonumber\\
&& + \sum_{k=1}^{\infty} (-1)^k e^{- \ve k+ i \pi s k } \cr
&=& \rme^{-i {\pi s/2 }}\frac{1}{1+ e^{-\ve - i \pi s}}+ \frac{1}{1+ e^{-\ve + i \pi s}}-1\cr
&=&\rme^{-i {\pi s/2 }}\frac{\sinh\ve}{\cosh{\ve}+\cos{\pi s}}\,,
\ea
yielding the desired result. The bosonic case is a variation on the same theme which presents no difficulty.


\subsection{Bosonic and susy solutions with wild oscillations: Comparison with the literature}\label{compwo}

A completely different way of approaching the problem was adopted in \cite{CST} for the bosonic case and subsequently developed in Ref.~\cite{FGN2}. In \cite{CST}, a level truncation analysis of the tachyon dynamics was carried out for a perturbative solution given as a finite sum of exponentials of the form
\begin{equation}\label{rollsum}
\phi(t)=\sum_{n=1}^{n_{\rm max}} a_n \rme^{n t}\,.
\end{equation}
The solution and all its derivatives satisfy the boundary condition $\phi^{(p)}\to 0$ as $t\to -\infty$. The first three coefficients are exact, since $a_n$ can be related to the (exact) $n+1$ scattering amplitude \cite{FGN2}. The remaining coefficients ($n\ge 4$) can be perturbatively obtained by imposing that the trial function \Eq{rollsum} satisfies the cubic equation of motion in the bosonic case at increasing levels. Reliable numerical values of $a_n$ were known only up to $a_6$ \cite{CST,FGN2}, but recently an analytic bosonic expression for $a_n$ has been derived \cite{KORZ}. 

For negative $t$, Eq.~\Eq{rollsum} with the appropriate coefficients $a_n$ describes the rolling of the tachyon off the unstable maximum along the potential. The physical interpretation for positive $t$ is more problematic. The
truncated expansion (\ref{rollsum}) is a solution only up to some upper bound $t=t_b$, which increases by increasing the number of terms one includes in the sum. Consequently, the asymptotic behaviour of the solution for large positive  $t$ cannot be extrapolated from Eq.~\Eq{rollsum}: being the sum alternate, $\phi\sim \pm \infty$ depending on the order $n$ at which one truncates the sum \Eq{rollsum}.

Before exploding exponentially, the field $\phi(t)$ presents an oscillatory behaviour with increasing amplitudes that makes the rolling tachyon dynamics difficult to interpret. In particular, the width of oscillations for $t>0$ is well beyond the classical inversion point on the tachyon potential, apparently violating conservation of the total energy.

\begin{figure}\begin{center}
\includegraphics[width=8.6cm]{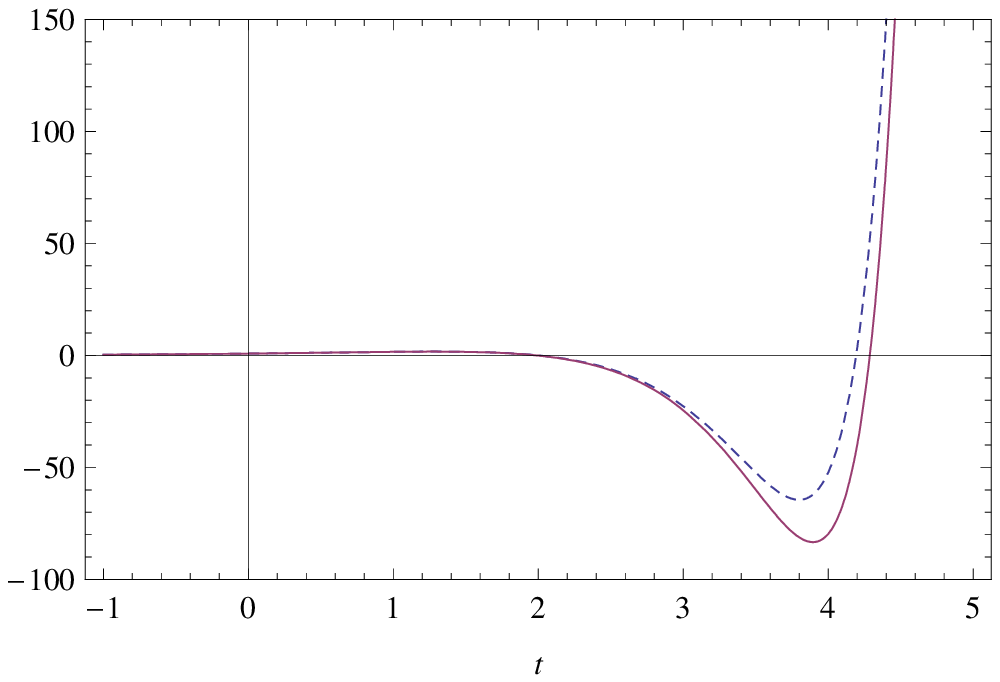}
\includegraphics[width=8.6cm]{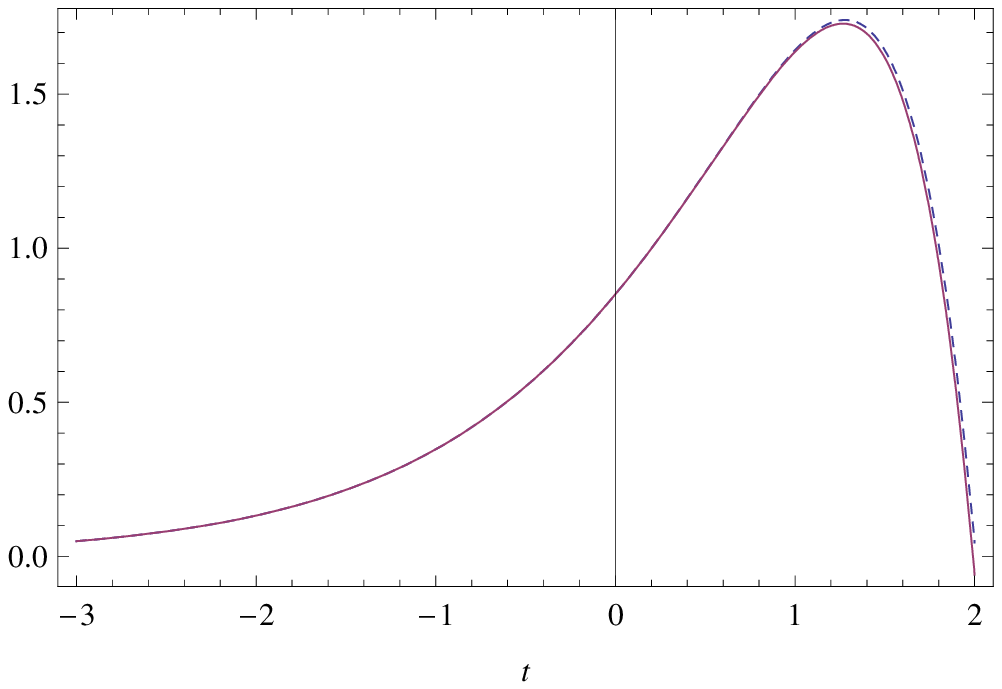}
\caption{\label{fig2} The nonperturbative [Eq.~\Eq{psix}, solid line] and perturbative (\cite{CST,FGN2}, dashed line) bosonic solutions with wild oscillations. The two curves are zoomed in in the second panel.}
\end{center}\end{figure}

For $n\leq 2$, the coefficients $a_n$ of Eq.~\Eq{rollsum} are identical to the ones defining our analytic solution Eq.~\Eq{psix}, even though the latter corresponds to a $(0,0)$ level truncation; for $n>2$ they are very close. In Fig.~\ref{fig2}, Eq.~\Eq{psix} is compared to the perturbative solution discussed in \cite{CST,FGN2}. The two curves are practically overlapped up to $t=2$. For $2<t<4$ there are some small deviations.

\begin{figure}\begin{center}
\includegraphics[width=8.6cm]{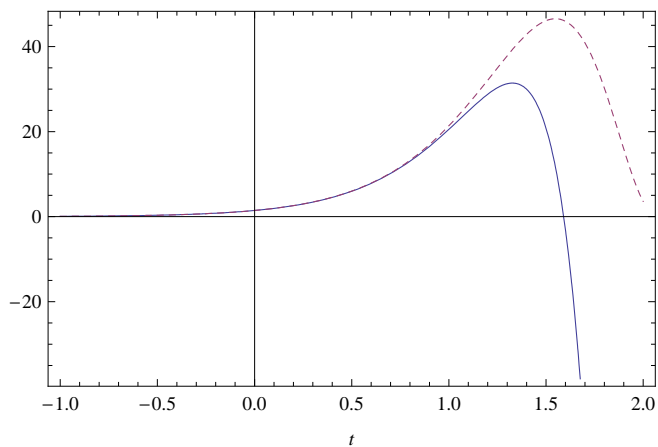}
\caption{\label{fig3} The left- and right-hand side of Eq.~\Eq{aref2} (solid and dashed curve, respectively) when the oscillating function Eq.~\Eq{nloc3x} is plugged in.}
\end{center}\end{figure}

While in the perturbative method the convergence abscissa $t_b$ of the solution is unknown (because the full tower of coefficients $a_n$ is unknown beyond the truncation point), here one can verify the equations of motion at any time (because they are localized). To check whether $\psi_-(r,t)$ is a solution of the SFT equation also for $t>0$, one can substitute it in Eq.~\Eq{aref2}. It turns out that this is the case, at least up to some $t_*=O(1)$ (see Fig.~\ref{fig3}).

The heat equation method allows one to find both the bounded even solution and another (being the analytic continuation of the left half of the former to the region $t>0$) with wild oscillations. The two classes of solutions seem, so far, not mutually exclusive, inasmuch as the choice of one instead of the other is dictated by different requirements. Analyticity at the origin selects the wildly oscillating solution, while boundedness and inversion of the nonlocal operator (the possibility to recover the local solution by applying $\rme^{-r_*\B}$ to the nonlocal solution) support the even solution.

Therefore, this oscillating solution is nonperturbative (i.e., an infinite convergent series with known coefficients) but limited at lowest truncation level in the string spectrum. On the other hand, the approach of \cite{CST,FGN2} constructs an oscillating perturbative series (that is, finite and with numerical coefficients) which is higher-level (in the string spectrum) and whose radius of convergence cannot be calculated but seems to be infinite.


\subsection{The physical picture: Choice of the oscillating solution}\label{phys}

We have at hand two possible approximate solutions of supersymmetric SFT: either Eq.~\Eq{nloc3} or Eq.~\Eq{nloc3x}. However, none of the two is completely satisfactory, both on mathematical and physical grounds. One solution is well-defined on the whole temporal axis but has a spike at the origin, where one must define its derivatives in a delicate way. The other solution does not have these problems, it is well-defined at the origin but can be found only up to some finite convergence abscissa and, for $t>0$, presents ever-growing oscillations.

Physically, the situation may seem even more obscure. We first summarize the properties of the corresponding local solution $\phi(0,t)={\rm sech}t$ ($\phi>0$ from now on). At $t=-\infty$ it is at rest at the unstable (perturbative) vacuum, that is, the local maximum of the potential $V$. As time passes, the field rolls down the potential and passes through the minimum, where the kinetic energy is maximal. Since the energy is conserved and the system is classical, the field cannot stop at the minimum and proceeds towards an inversion point $\phi^*$ defined, through energy conservation, by the condition $V(0)=V(\phi^*)$. This happens at $t=0$, where the field reaches the maximum value. For $t>0$, the tachyon inverts its motion, passing again through the minimum and reaching asymptotically the unstable maximum. The behaviour of $\phi(0,t)$ is clear as dictated by energy conservation.

For $t<0$, both candidate nonlocal solutions behave in a way similar to that of the local solution. For $t>0$, one solution suddenly bounces back before reaching the inversion point required by energy conservation of a canonical particle. This would happen if a rigid wall was placed between the minimum and the inversion point of the potential, while in this case the potential does not have any such feature. On the other hand, the second solution passes the inversion point and proceeds further along the potential, reaching energy levels that a canonical, local system would not have at $t=-\infty$.

To understand these facts, one has to abandon the misleading picture of a standard particle with a given kinetic energy moving in a potential. The point is that the self-interaction felt by the particle is not given by the potential. Rather, it is the potential dressed by the kinematic (nonlocal) factors $\rme^{-r \p^2_t}$. Their presence drastically changes the dynamics, as one can see by the following heuristic arguments valid on any homogeneous background. Let us consider slowly varying profiles. In this case, the interaction term Eq.~\Eq{pot} can be expanded as (we ignore the cosmological constant)
\ba
\tilde V&=&\frac{1}{2} m^2\phi^2 +\frac{\s}{n}\left(\rme^{r\B}\phi\right)^n\nonumber\\
&\approx&\frac{1}{2}m^2\phi^2 +\frac{\s}{n}\phi^n+ r\s\phi^{n-1} \B\phi\nonumber\\
&=& V+ r\s\phi^{n-1} \B\phi\,,\label{slow}
\ea
where $V$ is the potential of the local solution $\phi(0,t)$, and the additional term can be thought as a modification of the kinetic energy $E_{\rm kin}$. The Lagrangian is $\approx\phi\B\phi/2-r\s\phi^{n-1} \B\phi-V\to -(\nabla\phi)^2/2+(n-1)r\s\phi^{n-2}(\nabla\phi)^2-V$ via an integration by parts, and
\be
\label{kin}
E_{\rm kin}\approx \tfrac{1}{2}\dot \phi^2 \left[1-2(n-1)r\s\phi^{n-2}\right]\,.
\ee
During slow rolling, the major contribution due to the nonlocality of the potential affects the kinetic term rather than the potential. Specializing to the bosonic string and positive field values, if $r<0$ the square bracket in the right-hand side of Eq.~\Eq{kin} is positive definite ($n,\s>0$) and the usual physical interpretation of motion along the potential exchanging kinetic and potential energy holds. On the contrary, $r=0$ is a bifurcation point for the system, as for $r>0$ the square bracket can become negative. More precisely, if $r>0$ there is a critical value for the tachyonic field, $\phi_c(r)=1/(4r\sigma)$, at which $E_{\rm kin}\approx0$.

Two cases are possible. If any solution of the equation of motion exceeds $\phi_c$, then the effective kinetic energy contributes with a negative sign to the total energy, and conservation of the latter forces the tachyon to go indefinitely up the potential. This happens to the oscillating solution, which is related to the perturbative solution analyzed in \cite{CST,FGN2}. Or, since $\phi_c$ is located between the minimum of the potential and the local inversion point, if the tachyon solution is bounded it must be so by $\phi_c$ and not by the local inversion point, otherwise it would undergo the previous phenomenon. On the other hand, when the tachyon arrives at $\phi_c$ it has not exhausted its velocity, being below the inversion point. It cannot stop, otherwise it would violate energy conservation, and the only thing it can do compatibly with the latter is bouncing back rigidly. Then, the tachyon field has a spike and its velocity changes sign.\footnote{Another example of physical models with discontinuities is thermal systems displaying one or more discontinuous phase transitions. The relevant parameter there is the total energy, which is a constant observable. This is also our case, as one can verify numerically from Eq.~\Eq{energy}.} 

Within this simple approximation, all the main characteristic features of the solutions we discussed are recovered, including not only the spike and the indefinite growth but also the bifurcation of the solutions when $t>0$ in the $r>0$ case. One can check the validity of this qualitative picture, for instance, in the bosonic case, where
\be
\frac{\phi_c(r_*)}{\phi(r_*,0)}\approx0.83,
\ee
which misses the value of the spike by only 17\%.

As mentioned in the introduction, the tachyon with wild oscillations has been confirmed as a solution of the full equation of motion \cite{KORZ}. \emph{A posteriori}, this result is not surprising. At level $(0,0)$, Schnabl's gauge coincides with Siegel's gauge, and the effective equation of motion for the tachyon is the same. As the truncation level in the Siegel gauge increases, the shape of the effective tachyonic potential changes only in the quantitative details of the local minima. Hence, one would expect that the features of solutions at low levels would survive through the truncation procedure. This guess was confirmed in Ref.~\cite{KORZ} for the case of the wildly oscillating solution. Also, evidence was given that the radius of convergence of the series defining this solution is actually infinite for any $t$.
In order to complete the comparison, one can note that the analogous of the bounded solution in the Schnabl gauge for the exact equation of motion exists and is the mirroring at $t=0$ of the solution of \cite{KORZ} for negative times, as explained above.

In the diffusion equation method, the spike solution is more accurate than the oscillating one and has the advantage of admitting several integral representations. However, there are reasons to believe that the oscillating solution Eq.~\Eq{nloc3x} is actually the only feasible one:
\begin{enumerate}
\item[(i)] As said in Sec.~\ref{thesol}, the spiky solution emerges because we fixed the asymptotes at both the infinite past and future. In other words, we have imposed that the origin continues to be the inversion point of the solution even in the nonlocal system. However, nothing forces us to do so and, rather, the most natural option is to leave the solution unconstrained asymptotically.
\item[(ii)] The only solution found perturbatively is the oscillating one. The point $t=0$, where the two solutions bifurcate, is not special in the perturbative approach.
\item[(iii)] When two solutions overlap before a bifurcation point which is not physically special, they are physically equivalent under a gauge transformation. In the case under scrutiny there is no such gauge freedom, so one solution must be discarded.
\end{enumerate}
These arguments, together with others in Sec.~\ref{rsbbs}, show that one can safely consider the solution with wild oscillations as the only meaningful alternative. This will be our attitude in what follows. Before proceeding, it may be interesting to remark that one of the most common objections against the spike (``The spiky solution rolls back to the local maximum and this certainly does not describe tachyon condensation at the minimum'') cannot be moved to exclude it. In fact, the same can be said regarding the oscillating solution, which does not sit down to the local minimum but climbs the potential indefinitely back and forth.

For heuristic reasons, in this section we have tried to give a partial explanation of the wild oscillations in terms of the effective change of sign of an effective kinetic energy. However, it was shown in \cite{CST,ell07} that these oscillations disappear after a suitable gauge transformation interpolating between OSFT and BSFT. In our approach (effective OSFT lowest-level tachyonic action) it is not possible to see this mechanism in action, the gauge being fixed from the beginning; hence we cannot establish whether the oscillations of the solution are physical or gauge artifacts. However, we shall derive in Sec.~\ref{bccor} a relation between OSFT and BSFT governed by the sign of the parameter $r$, supporting the idea that the framework of BSFT has more transparent physical features.


\section{Bosonic BSFT}\label{bsftsol}

\subsection{General setup}

In Witten's construction of open boundary string field theory~\cite{wit92}, the space of all two-dimensional worldsheet field theories on the unit disk, which are conformal in the interior of the disk but have arbitrary boundary interactions, is described by the worldsheet action 
\begin{equation}
\mathcal{S}=\mathcal{S}_0+\mathcal{S}_{\rm boundary}=
\mathcal{S}_0+\int_0^{2\pi}\frac{\rmd\tau}{2\pi}\,\mathcal{V}\,. \label{Sboun}
\end{equation}
Here, $\mathcal{S}_0$ is the bulk action, a free action describing an open plus closed conformal background integrated over the volume of the unit disk, and $\mathcal{V}$ is a general perturbation defined on the disk boundary which can be parametrized by couplings $\lambda^i$,
\begin{equation}
\mathcal{V}=\sum_i \lambda^i \mathcal{V}_i\,.
\label{nu}
\end{equation}
The couplings $\lambda^i$ correspond to fields in spacetime, and according to~ \cite{wi92b,sh93a} the classical spacetime action $S$ is defined by
\begin{equation}
S = \left(\sum_i\beta^i\frac{\partial}{\partial\lambda^i}+1\right)Z\,,\label{act}
\end{equation}
where $Z$ is the disk partition function of the worldsheet theory (\ref{Sboun}) and $\beta^i$ are the $\beta$ functions of the couplings governing their worldsheet renormalization group flow. 
For open strings propagating in a tachyon background the worldsheet action (\ref{Sboun}) reads
\ba
\mathcal{S}[X]&=&\int \rmd\sigma \rmd\tau \frac{1}{4\pi}\partial_a X(\sigma,\tau)\cdot \partial^a X(\sigma,\tau)\nonumber\\
&&+\int_0^{2\pi}\frac{\rmd\tau}{2\pi} T[X(\tau)]\,,\label{actionint}
\ea
and the partition function
\begin{equation}
Z=\int [\rmd X]\,\rme^{-\mathcal{S}[X]}\,.
\label{partf}
\end{equation}
Via a standard procedure~\cite{CT1,CT2,CT3} the bulk excitations can be integrated out to get an effective field theory~\cite{GLOS} which lives on the boundary~\cite{CFGNO}:
\begin{equation}
Z(J)=\int [\rmd X]\,\rme^{-\int_0^{2\pi}\frac{\rmd\tau}{2\pi}
\left[\frac{1}{2}X^\mu |\rmi\partial_\tau | X_\mu + T(X)-J\cdot X\right]}\,,
\label{zj}
\end{equation}
where $\cdot$ denotes the scalar product of Lorentz vectors, $J_\mu(\tau)$ is the usual source generating correlators of the fields $X^\mu$ restricted to the boundary of the worldsheet, and the operator $|\rmi\p_\tau|$ is defined by the Fourier series
\be
|\rmi\partial_\tau|\,\delta(\tau-\tau')=\sum_{n=-\infty}^{+\infty} \frac{|n|}{2\pi}\,\rme^{\rmi n(\tau-\tau')}\,.
\ee
$Z(J)$ in Eq.~(\ref{zj}) is defined up to a multiplicative constant $c$ which, in turn, is just the tension of the $D25$-brane~\cite{CFGNO,KMM2}.

To calculate the energy, we split $X^\mu$ in a classical term $x^\mu$ (constant in worldsheet coordinates) and a varying part (which we still call $X^\mu$), promote the Minkowskian metric in Eq.~\Eq{partf} to a general one, $\eta^{\mu\nu}\to g^{\mu\nu}$, and use the standard definition of the energy-momentum tensor. The spacetime action is proportional to the partition function (via a positive constant $c$) and one has \cite{se021,lar02}
\be
T_{\mu\nu} \equiv -\frac{2c}{\sqrt{-g}}\frac{\delta Z}{\delta g^{\mu\nu}} = c(g_{\mu\nu}Z+{\cal A}_{\mu\nu})\,,\label{emt}
\ee
where
\ba
{\cal A}_{\mu\nu} &\equiv& 2\int [\rmd X]\,\left(\int \rmd\sigma \rmd\tau \frac{1}{4\pi}\partial_a X_\mu \partial^a X_\nu\right)\,\rme^{-\mathcal{S}[X]}\nonumber\\
&=& 2\int [\rmd X]\,\,\partial_a X_\mu(0) \partial^a X_\nu(0)\,\,\rme^{-\mathcal{S}[X]}\,.\label{emtA}
\ea
The first term in Eq.~\Eq{emt} comes from the variation of the extra factor $\rmd^D x\,\sqrt{-g}$ which appears in the measure of the integral in $X^\mu$; the zero mode has been integrated out in the partition function $Z$. The second term corresponds to the expectation value of the graviton vertex operator and is found under the assumption that the boundary interaction is independent from the metric. In the last line, the position of this operator was fixed.


\subsection{Rolling solution of bosonic boundary SFT}\label{rsbbs}

If we consider the case of  constant source  $\rmi k_\mu$ for the zero mode of the $X^\mu$ field, the integral over the zero mode variable will just provide the energy-momentum conservation $\delta$ function. In this case, the partition function (\ref{zj}) becomes
\begin{equation}
Z(k)=\int [\rmd X]\,\rme^{-\int_0^{2\pi}\frac{\rmd\tau}{2\pi}
\left[\frac{1}{2}X^\mu |\rmi\partial_\tau| X_\mu + T(X)\right]-\rmi k\cdot x}\,,
\label{zetak}
\end{equation}
where $x$ is the zero mode defined by
\be
x^\mu = \int_0^{2\pi}\frac{\rmd\tau}{2\pi} X^\mu(\tau)\ .
\label{zerom}
\ee
In order to evaluate the path integral (\ref{zetak}), we need the Green function $G$ of the operator $|\rmi\partial_\tau|$, $|\rmi\partial_\tau|G(\tau')=\delta(\tau-\tau')$,
\be
G(\tau)=2\sum_{n=1}^{\infty}\rme^{-\ve n} \frac{\cos n\tau}{n}=
-\ln \left[1-2\rme^{-\ve}\cos \tau +\rme^{-2\ve}\right],\label{prop}
\ee
where $\ve$ is an ultraviolet cutoff. Clearly, $G(\tau)$ is defined up to 
an arbitrary constant, which is the kernel of the operator $|\rmi\partial_\tau|$.
Regularizing the propagator as in Eq.~(\ref{prop}) and adding the arbitrary constant 
parametrized as $2r$, we are led to the following prescription for $G(\tau)$:
\ba
G(\tau)=\begin{cases}
-\ln\left[4\sin^2\left(\frac{\tau}{2}\right)\right]+2r & \tau\ne 0 \cr
-2\ln\ve & \tau=0\end{cases}\,.
\label{prop0}
\ea
In the second line the arbitrary constant $r$ has been absorbed in the regulator $\ve$. The parameter $r$ is an ambiguity stemming from different prescriptions in the scheme adopted to renormalize the effective action \cite{tse00,GLOS}. As this affects vertex operators, it corresponds to a field redefinition of target fields. In Appendix \ref{bounda} it will be shown that $r$ naturally arises also in a canonical quantization framework, and it will be related to the ordering prescription in the evaluation of quantum correlators. The usual normal ordering corresponds to the choice $r=0$.
 
The partition function (\ref{zetak}) can be evaluated as an expansion in powers of the bare fields $T[X(\tau)]$. Taking 
the Fourier transform of the tachyon and performing all the contractions of the $X(\tau_i)$ fields, we get~\cite{CFGNO}
\ba
Z(k)&=&\sum_{n=0}^\infty\frac{(-1)^n}{n!}\,\ve^{-n}\int 
\prod_{i=1}^{n}\rmd k_i\, T(k_i)\int_0^{2\pi}\prod_{i=1}^{n}\frac{\rmd\tau_i}{2\pi}\nonumber\\
&&\times\rme^{-\sum_{i=1}^n\frac{k_i^2}{2}G(0)-\sum_{j>i} k_i\cdot k_j G\left(\tau_i-\tau_j\right)}\nonumber\\
&&\times\delta\left(k-\sum_{i=1}^n k_i\right),
\label{zetan}
\ea
where we have omitted the vector indices in the $\delta$ function. Taking into account the propagator (\ref{prop0}) and evaluating the integrand on the support of the $\delta$ function, we obtain
\ba
Z(k)&=& \rme^{-r k^2} \hat Z (k)\nonumber\\
&=& \rme^{-r k^2}\sum_{n=0}^\infty\frac{(-1)^n}{n!}\int 
\prod_{i=1}^{n}\rmd k_i\, \hat{T}(k_i)\,\delta\left(k-\sum_{i=1}^nk_i\right) \cr &&\times\int_0^{2\pi}\prod_{i=1}^{n}\frac{\rmd\tau_i}{2\pi}
\prod_{j>i}\left[4\sin^2\left(\frac{\tau_i-\tau_j}{2}\right)\right]^{k_i\cdot k_j},
\label{zetaq}
\ea
where $\hat{T}(k_i)\equiv T(k_i) \ve^{k^2_i-1}\rme^{r k^2_i}$. Apart from a trivial rescaling of the tachyon fields, all the $r$-dependence in $Z(k)$ can be factorized out of the integrals by an overall factor $\rme^{-r k^2}$. In fact, $\hat Z(k)$ is nothing but the partition function for the tachyon field $\hat T$ when $r=0$. Rolling tachyon solutions can be obtained by the following choice of the bare tachyon fields
\ba
T(X)&=& T_0\,\rme^{\rmi p\cdot X},\nonumber\\
T(k_i)&=&\frac{1}{(2\pi)^D}\int \rmd X \rme^{-\rmi k_i\cdot X} T(X)=T_0 \delta (k_i - p)\,.\nonumber\\
\label{T}
\ea
This corresponds to the case in which all the momenta $k_i^\mu$ in Eq.~(\ref{zetaq}) have the same value $p^\mu$ (coherent phases). Such a profile is particularly simple and corresponds, in the Minkowskian formulation, to a perturbation around the unstable vacuum at $X^0=-\infty$ if the momenta $k_i^\mu$ are purely time-like (in that case, $T_0$ is the tachyon velocity at $X^0=0$). Moreover, the integrals over $\tau_i$ can be now explicitly performed by using the formula
\be
\int_0^{2\pi}\prod_{i=1}^{n}\left(\frac{\rmd\tau_i}{2\pi}\right) 
\prod_{j>i}\left[2 \sin\left(\frac{\tau_i-\tau_j}{2}\right)\right]^{2p^2}=
\frac{\Gamma (1+ n p^2)}{[\Gamma(1 + p^2)]^n}\,.
\label{rama}
\ee
To get the partition function $Z(X)$ in the coordinate space we have to Fourier transform Eq.~(\ref{zetaq}). Taking Eq.~(\ref{rama}) into account we obtain
\ba
Z(X) &=&\rme^{-r\Box}\sum_{n=0}^\infty \frac{(-1)^n}{n!}\left[\TT(X)\right]^n\Gamma(1+n p^2)\,,\nonumber\\
\TT(X)&=&\ve^{p^2-1}\frac{T_0 \rme^{\rmi p\cdot X}\rme^{rp^2}}{\Gamma(1+p^2)}\,.
\label{zq1}
\ea
The sum over $n$ can be perfomed if the Euler representation for $\Gamma(1 + n p^2)$ is used. We get
\be
Z=\rme^{-r\Box}\int_0^\infty \rmd s\, \rme^{-s[1+\TT(X)s^{p^2-1}]}\,.
\label{zq2}
\ee
The renormalized tachyon field $\vp (X)$ is related to the partition function $Z(X)$ by the formula~\cite{CFGNO} 
\be 
Z(X)=1- \vp(X)\,.\label{zt}
\ee
The equation of motion for the renormalized field are obtained by imposing the vanishing of the corresponding $\beta$ function
\be
\beta_{\vp}\equiv - \frac{\partial \vp}{\partial \ln \ve}=0\,,
\label{beta}
\ee
which enforces the condition $p^2=1$, as one can verify. We shall consider the Wick rotated back profile
($\rmi X^0= t$, $-\rmi p_0=p^{\rm Eucl}_0$) in the spatially homogeneous case, $p=(1,0,\dots ,0)$ ($p^{\rm Eucl}_0=-\rmi$). The integration over $s$ in Eq.~(\ref{zq2}) is now trivial, but the effect of the (Wick rotated) operator $\rme^{-r \partial_t^2}$ on it is cumbersome. It is preferable to first rewrite the integral in
(\ref{zq2}) as
\be
\int_0^\infty \rmd s\, \rme^{-s[1+\mathcal T (t)]}=
-\frac{1}{2 \rmi}\int_{\Gamma_R}\rmd \s\, \frac{[\TT (t)]^\s}{\sin(\pi \s)}\,,
\label{gammar}
\ee
where $\TT(t)=T_0 \rme^{t+r}$, and the contour $\Gamma_R$ lies on the imaginary $\s$ axis from $-\rmi \infty$ to $+\rmi \infty$ keeping the pole in $\s=0$ to its right. Then, from Eq.~\Eq{zt} the renormalized tachyon reads
\be
\vp(r,t)=\frac{1}{2 \rmi} \rme^{-r\p_t^2}\int_{\Gamma_L}\rmd \s \frac{[\TT(t)]^\s}{\sin(\pi \s)}\,,
\label{gammal}
\ee
where we have written the dependence on $r$ as an argument of $\vp$ and the contour $\Gamma_L$ now keeps the pole in $\s=0$ to the left. When $r=0$ one should recover the case analyzed in Ref.~\cite{lar02}. Indeed, without the operator $\rme^{-r\p_t^2}$ one gets
\ba
\vp_+(0,y)=\sum_{n=0}^\infty(-1)^n\, \rme^{-ny},\;\;\;\;
\;\; y >0\,,\cr
\vp_-(0,y)=-\sum_{n=1}^\infty (-1)^n\, \rme^{ny},\;\;\;\;
\;\; y<0\,,
\label{q1}
\ea
where $y\equiv t+\ln T_0$. These expressions are obtained as the sum over the residues of the function $1/\sin \pi \s$ closing the integral over $\s$ to the left or to the right depending on the sign of $y$. Both the expressions $\vp^{(\pm)}(0,y)$ in (\ref{q1}) sum to
\be
\vp(0,y)=\frac{1}{1+ \rme^{-y}}=\frac{1}{2}+\frac{1}{2} \frac{\sinh y}{\cosh y+1}\,.
\label{tanh}
\ee
This expression exactly represents the well-known rolling tachyon solution of bosonic BSFT~\cite{lar02}, as can be easily seen by recalling the definition of the tachyon effective potential in terms of the renormalized field $1-\vp=\rme^{-\tilde T}$ \cite{CFGNO}:
\be
U = \left(1-\vp\right)\left[1-\ln \left(1-\vp\right)\right]
= \rme^{-\tilde T}(1+\tilde T)\,.\label{BSFTtachpot}
\ee
At the infinite past $t\to -\infty$, the tachyon $\tilde T=0$ starts from the unstable maximum of the potential (\ref{BSFTtachpot}), reaching the stable vacuum at $t\to+\infty$ as $\tilde T\to +\infty$. Expanding Eq.~\Eq{BSFTtachpot} for small $\tilde T$, one reproduces the cubic potential. From Eqs.~(\ref{zt}) and (\ref{tanh}), one has precisely the partition function of Ref.~\cite{lar02}
\be
Z(r=0)=\frac{1}{1+T_0\rme^t}\,.
\label{larsen}
\ee 
The value of the partition function (as well as $\vp$) at $y=0$ is defined only by analytic continuation.

A solution for $r<0$ can be obtained directly from Eq.~(\ref{gammal}) by changing variable $\s \to -\rmi \s$, leading to 
\be
\vp(r,t)=\frac{1}{2}+\frac{1}{2}\int_{-\infty}^{+\infty} \rmd\s\, \rme^{r\s^2} \frac{\sin
(\s y)}{\sinh (\pi\s)}\,,\qquad r<0\,,
\label{trint}
\ee 
where $y\equiv t+r+\ln T_0$. The factor $1/2$ comes from the integral over the half-circle around the origin. The profile in Eq.~(\ref{trint}) is $C^\infty$ and generalizes the solution (\ref{tanh}) and its rolling behaviour for any $r\le 0$ [for $r=0$ it coincides with Eq.~(\ref{tanh})].

In order to get a solution in the region $r>0$, one can start from Eq.~\Eq{tanh} and apply the method of Appendix \ref{bsrel}, or by analytic continuation of the series representation. Both calculations give the same result:
\be
\vp(r,t)=\frac{1}{2}+\frac{1}{2}\int_{-\infty}^\infty \rmd\s\, K(\s,r)\,\frac{\sinh y}{\rme^\ve \cosh y + \cos\s}\,.\label{trreg}
\ee
The behaviour at the origin $y=0$ in the $r>0$ case is regulated by a mechanism analogous to that described in the OSFT case and we shall not repeat it here. The solution $\vp$ in any of its representations [for instance (\ref{trint}) or (\ref{trreg})] satisfies the diffusion equation with $-1$ diffusion coefficient
\be 
\p_r\vp(r,t)=-\p_t^2\vp(r,t)\,,
\label{diffusion}
\ee
with respect to the ``radial'' variable $r$ and the time variable $t$. In fact, $\vp(r,t)$ is nothing but the solutions of this diffusion equation with ``initial'' condition Eq.~\Eq{tanh} and ``boundary'' condition $\vp(r, \pm \infty)=1/2\pm 1/2$. The effect of $r$ in the rolling solutions is twofold: it translates the origin of time and it changes the slope of the rolling. The first effect can be always reabsorbed by a suitable time translation, under which the system is invariant; in alternative, one fixes $T_0=\rme^{-r}$.

\begin{figure}\begin{center}
\includegraphics[width=8.6cm]{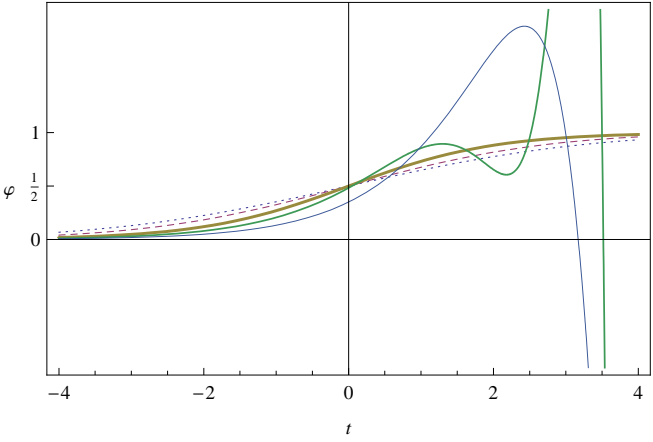}
\caption{\label{fig4} The BSFT solution for different values of $r$, given by Eq.~\Eq{trint} ($r\leq 0$, kink-type) and \Eq{q2box} ($r>0$, wild oscillations). In the panel, $r=-2$ (dashed curve), $r=-1$ (dotted curve), $r=0,\,0.5,\,1$ (solid curves with decreasing thickness).}
\end{center}\end{figure}

In Appendix \ref{bsrel} we discuss the series representation of the solution \Eq{trreg} for $r>0$, which is
\ba
\vp_+(r,t)=\sum_{n=0}^\infty (-1)^n \rme^{-rn^2} \rme^{-n y},\;\;\;\;
\;\;y>0\; ,\ \ \ r>0\,, \cr
\vp_-(r,t)=-\sum_{n=1}^\infty (-1)^n \rme^{-rn^2} \rme^{n y},\;\;\;\;
\;\; y<0\; ,\ \ \ r>0\,.\cr
\label{q2bo}
\ea
There, we show that the series and integral representation are exactly equivalent, and they are the analytic continuation of the solution with $r<0$. As in the OSFT case, one can also consider the solution 
\be
\vp_-(r,t)=-\sum_{n=1}^\infty (-1)^n \rme^{-rn^2} \rme^{n y},\label{q2box}
\ee
at all times; its integral representation can be found with the same technique of Sec.~\ref{thesol}. The behaviour of Eq.~\Eq{q2box} for different values of $r$ is shown in Fig.~\ref{fig4}. When $r>0$ the solution oscillates at positive times.

Any evaluation of the path integral \Eq{zj} based on the Taylor expansion in powers of the bare fields $T[X(\tau)]$ unavoidably leads to Eq.~\Eq{q2box}, which is the representation of $\vp$ [or $Z$, through Eq.~\Eq{zt}] as a power series of $T$. An alternative route is through the boundary states~\cite{se021,se022}. Even in that case, one gets the power series Eq.~\Eq{q2box}, as the calculation is expressed in terms of correlators, and therefore intrinsically perturbative. As the boundary state calculation does not show any sign of the decomposition Eq.~\Eq{q2bo}, we select Eq.~\Eq{q2box} as the correct family of BSFT solutions. Another reason is that Eq.~\Eq{q2bo} would lead to inconsistencies in its interpretation as an instantonic particle, since the latter could cover finite lengths instantaneously.


\subsection{Energy--momentum tensor}

It is convenient to define the normal-ordered graviton vertex operators
\ba
:\p X^\mu(z,\bar z)\bar\p' X^\nu(z',\bar z'): &\equiv& \p X^\mu(z,\bar z)\bar\p' X^\nu(z',\bar z')\nonumber\\
&& +g^{\mu\nu} \p\bar\p'\ln\rme^{-r} |z-z'|\,,\nonumber\\
\ea
and
\ba
\ono\p X^\mu (z, \bar z)\bar\p' X^\nu (z' ,\bar z') \ono &\equiv&  \p X^\mu(z,\bar z)\bar\p' X^\nu(z',\bar z')\nonumber\\
&& +g^{\mu\nu} \p\bar\p'\ln\rme^{-2r} |z-z'|^2\nonumber\\
&=& :\p X^\mu \bar\p' X^\nu :-\frac{g^{\mu\nu}}{2}
\,,
\ea
where $\p=\p_z$ and $\bar\p=\p_{\bar z}$. In agreement with our definition of the propagator, we have generalized the expressions of \cite{lar02} for $r\neq 0$. This operation is clearly trivial, and the calculation of \cite{lar02} (to which we refer the reader for intermediate steps) is reproduced. Setting $g^{\mu\nu}=\eta^{\mu\nu}$, one has ${\cal A}_{ij}=\delta_{ij}Z$, while
\ba
{\cal A}^{00} &=& 2\langle \ono\p X^0(0)\bar\p X^0(0)\ono\rme^{-\int\frac{\rmd\tau}{2\pi} T[X^0(\tau)]} \rangle -Z\nonumber\\
&=& 2\sum_{n=0}^{+\infty} \frac{(-T_0 \rme^t)^n}{n!}\nonumber\\
&&\times\langle \ono\p X^0(0)\bar\p X^0(0)\ono 
\prod_{i=1}^n \int\frac{\rmd\tau}{2\pi}\rme^{X^0(\tau)} \rangle -Z\nonumber\\
&=& 2\sum_{n=1}^{+\infty} \frac{(-T_0 \rme^t)^n}{n!}\,\rme^{-2 rn(n-1)/2}n! -Z\nonumber\\
&=& 2\sum_{n=1}^{+\infty} (-\rme^{y})^n\,\rme^{-rn^2} -Z\nonumber\\
&=& -(1+\vp)\,.
\ea
Combining this expression with Eq.~\Eq{emt}, the pressure and energy read
\be
p \equiv T_{11} = 2c(1-\vp)\,,\qquad E \equiv -T_{00} = 2c\,.
\ee
The energy is constant, as it should be in Minkowski, and the tachyon tends asymptotically to pressureless matter.


\section{Supersymmetric BSFT}\label{sbsftsol}

One can extend the discussion of Sec.~\ref{bsftsol} to the case of superstrings. Typical unstable configurations where the open string contains a tachyon are non-BPS (Bogomolnyi--Prasad--Sommerfield) $Dp$-branes, with even $p$ for type IIB and odd $p$ for type IIA. This situation can be described through a perturbation of the worldsheet field theory by a boundary superpotential~\cite{wit98,HKM}
\begin{equation}
\mathcal{S}_{{\rm boundary}}=\int\frac{\rmd\tau}{2\pi}\int\rmd\theta\left[{\hat \zeta}D{\hat \zeta}+{\hat \zeta}T({\hat
X})\right]\,.\label{superpot}
\end{equation}
Here, a one-dimensional superfield notation is used to define worldsheet supercoordinates on the boundary,
\begin{equation}
{\hat X}^\mu (\tau,\theta)= X^\mu(\tau) + \theta \psi^\mu(\tau)\,, \label{supercoord}
\end{equation}
where $\theta$ is a Grassmann variable and $\psi^\mu$ is a Majorana fermion (in Neveu--Schwarz--Ramond formalism). 
In Eq.~(\ref{superpot}), $D=\partial_\theta+ \theta \partial_\tau$ is the derivative in superspace and the superfields ${\hat \zeta}$ are auxiliary anticommuting degrees 
of freedom encoding the Chan--Paton indices of the brane~\cite{wit98,HKM},
\begin{equation}
{\hat \zeta}^{I} (\tau,\theta) = \eta^I (\tau)+ \theta F^I(\tau)\,, \label{boundaryfermions}
\end{equation}
where $\eta^I$ is a propagating boundary fermion and  $F^I$ is an auxiliary field.
As in~\cite{lar02}, we will consider a single non-BPS $D$-brane. This implies the existence of
a single boundary fermion, and the index $I$ can be omitted. The superstring generalization
of the partition function (\ref{zj}) reads then
\begin{equation}
Z=P \int [\rmd{\hat X}][\rmd {\hat \zeta}] \rme^{-\mathcal{S}_0[{\hat X}]-\int\frac{\rmd\tau}{2\pi}\int
\rmd\theta\left[{\hat \zeta}D{\hat \zeta}+{\hat \zeta}T({\hat
X})\right]},\label{superzj}
\end{equation}
where  $P$
is the standard path-ordering operator, here nontrivial because of the tachyon-to-fermions coupling. In Eq.~(\ref{superzj}), we have ignored the presence of contact terms proportional to $T^2(\hat{X})$ \cite{se023,lar02}.

As in the bosonic case, the partition function can be evaluated perturbatively by expanding in powers of
the tachyon field. However, it is difficult to extract the $\rme^\B$ operator from $Z$ in momentum space, the reason being that one has to solve an integral much more involved than Eq.~\Eq{rama}. It is more convenient to adopt the  background field method~\cite{CFGNO,KS}. In this case one expands the fields ${\hat X}$ around a classical
background ${\hat X^\mu}(\tau,\theta)=x^\mu+{\hat Y^\mu (\tau,\theta)}$, where
$x^\mu$ satisfies the equations of motion and  varies slowly compared to the cutoff scale.
The supersymmetric version of the homogeneous time-dependent tachyon field (\ref{T}) can
then be written as
$$
T({\hat X})=\rme^{\frac{\rmi}{\sqrt{2}}x^0+ \frac{\rmi}{\sqrt{2}}\hat Y^0(\tau,\,\theta)}
\equiv \tilde{T} ~\rme^{\frac{\rmi}{\sqrt{2}} {\hat Y^0 (\tau,\,\theta)}},
$$
where $\tilde{T}=\rme^{\frac{\rmi}{\sqrt{2}}x^0}$ (we will neglect the tilde from now on). The partition
function reads as the following functional integral over the nonzero modes:
\ba
&&Z=\sum_{n=0}^\infty\frac{(-1)^n}{n!}\left(\frac{T}{\ve}\right)^n\int_0^{2\pi}
\prod_{i=1}^n\frac{\rmd\tau_i}{2\pi}\int\prod_{i=1}^n \rmd\theta_i\nonumber\\
&&\qquad\times \langle
{\hat\zeta}(\tau_1,\theta_1)\rme^{\frac{\rmi}{\sqrt{2}}{\hat Y}(\tau_1,\theta_1)} \cdots {\hat
\zeta}(\tau_n,\theta_n)\rme^{\frac{\rmi}{\sqrt{2}}{\hat Y}(\tau_n,\theta_n)}\rangle.\cr&&
\label{superz2} 
\ea
In order to evaluate this path integral, we need the supersymmetric extension of the Green function (\ref{prop}) that includes the fermionic two-point function on the disc (for a detailed derivation see~\cite{AT}):
\ba
{\hat G}_{ij}&=& 2\sum_{n=1}^{\infty}\rme^{-\ve n} \frac{\cos n\left(\tau_i-\tau_j\right)}{n}
 -2 \theta_i \theta_j\nonumber\\
 &&\times\sum_{q=0}^{\infty}\rme^{-\ve q} \sin \left[(q+1/2)\left(\tau_i-\tau_j\right)\right]\,.
\label{propsusy}
\ea
Adding the arbitrary constant consistently with (\ref{prop0}), we are led to the following definition for the Green function:
\ba
{\hat G}_{ij}&=& \langle {\hat Y}(\tau_i,\theta_i),{\hat Y}(\tau_j,\theta_j)\rangle\cr
 &=&-\ln \rme^{-2r}|z_i-z_j|^2-2\rmi\frac{\sqrt{z_i z_j}}{z_i-z_j}~\theta_i \theta_j\cr
 &=&-2\ln \rme^{-r}|z_i-z_j+\rmi\sqrt{z_i z_j}~\theta_i \theta_j|,~~~i\neq j,
 \label{SusyGreenFns}
\ea
where $z_i=\rme^{\rmi\tau_i}$ and we used the fact that $\{\theta_i,\theta_j\}=0$, $\theta_i\in \mathbb{R}$. The zero-point Green function ${\hat G}(0)\equiv G(0)$ is still defined through Eq.~(\ref{prop0}). The two-point function for the boundary fermions is defined by
\ba
\langle {\hat \zeta}(\tau_i,\theta_i),{\hat \zeta}(\tau_j,\theta_j)\rangle&=&{\hat
\Theta}_{i,j}\equiv {\hat\Theta}(\tau_i-\tau_j+\theta_i \theta_j)\cr
&=&\Theta(\tau_i-\tau_j)+\delta(\tau_i-\tau_j)~\theta_i \theta_j\,,\nonumber\\
\label{propferm}
\ea
where $\Theta(\tau_i-\tau_j)$ is the Heaviside step function. The partition function (\ref{superz2}) can now be formally written as
\ba
Z&=&\sum_{n=0}^\infty(-1)^n \left(\frac{T^2}{\ve}\right)^{n}\int_0^{2\pi}
\prod_{i=1}^{2n}\frac{\rmd\tau_i}{2\pi}\nn\\
&&\times\int\prod_{i=1}^{2n} \rmd\theta_i~ {\hat
\Theta}_{1,2}\dots {\hat
\Theta}_{2n-1,2n}\rme^{-\sum_{i=1}^{2n}\frac{G(0)}{4}-\sum_{j>i}\frac{{\hat
G}_{ij}}{2}}\cr
&=&\sum_{n=0}^\infty(-1)^n T^{2n}\int_0^{2\pi}
\prod_{i=1}^{2n}\frac{\rmd\tau_i}{2\pi}\nn\\
&&\int\prod_{i=1}^{2n} \rmd\theta_i~ {\hat
\Theta}_{1,2}\dots {\hat \Theta}_{2n-1,2n}\nn\\
&&\times\prod_{j>i}\rme^{-r}|z_i-z_j+\rmi\sqrt{z_i
z_j}~\theta_i \theta_j|\,, \label{superz3}
\ea
where the restriction to an even number of fields arises from the overall trace over the Chan--Paton
matrices here represented by the boundary fermions. Using the relation
\begin{equation}
|z_i-z_j+\rmi\sqrt{z_i z_j}~\theta_i \theta_j|=|z_i-z_j|+{\rm sgn}(\tau_i-
\tau_j)~\theta_i \theta_j\,,
\end{equation}
the integrals in (\ref{superz3}) can be evaluated as~\cite{lar02} 
\ba
\int_0^{2\pi} \prod_{i=1}^{2n}\frac{\rmd\tau_i}{2\pi}\int\prod_{i=1}^{2n} \rmd\theta_i~ {\hat
\Theta}_{1,2}\dots {\hat \Theta}_{2n-1,2n}\nn\\
\qquad\times\prod_{j>i}|z_i-z_j+\rmi\sqrt{z_i
z_j}~\theta_i \theta_j|=\frac{1}{2^n}\,.
\ea
After a Wick rotation $\rmi x^0=t$, the tachyonic profile is $T=\rme ^{t/\sqrt{2}}$ and the final result for the field Eq.~\Eq{zt} is then
\be
\vp_-=-\sum_{n=1}^\infty (-1)^n\rme^{-2rn^2}\rme^{ny}\,,\label{superz4}
\ee
where $y\equiv\sqrt{2}t+r-\ln 2$. This is a $q$ series with infinite radius of convergence for $r>0$ and valid for $y<0$ [see Eq.~\Eq{superz2}]. 

The same calculation of Appendix \ref{bsrel} yields the integral representation of the solution for negative values of $r$:
\begin{equation}
\vp=\frac{1}{2}+\frac{1}{2}\int_{-\infty}^{+\infty} \rmd\sigma\,\rme^{2 r \sigma^2}\,\frac{\sin(\sigma y)}{\sinh(\pi\sigma)}\,.\label{superz6}
\end{equation}
Modulo rescalings, Eq.~\Eq{superz6} is identical to the bosonic solution, Eq.~\Eq{trint}, and satisfies the diffusion equation $\B\vp=\partial_r\vp$.


\section{BSFT/OSFT correspondence and properties of rolling solutions}\label{bccor}

We are now ready to establish a correspondence between the bosonic OSFT and BSFT tachyon solutions we found. The relation between wildly oscillating solutions in series representation is clear from Eqs.~\Eq{psix} and \Eq{ttx}:
\be\label{bcc}
6 \dot\vp =\psi\,.
\ee
This relation can be verified also in integral representation. For simplicity we shall consider $t<0$, where the spiky and the wild oscillating solutions overlap, thus implying that Eq.~\Eq{bcc} holds for any solution and for any sign and value of the parameter $r$.

One starts with two real-valued functions
\be
u(\s,t) = \frac{\sin \s}{\cosh t + \cos\s}\,,\qquad v(\s,t) = \frac{\sinh t}{\cosh t + \cos\s}\,,\label{bosv}
\ee
which are harmonic conjugate,
\be
\p_t u+\p_\s v=0\,,\qquad \p_\s u-\p_t v=0\,.\label{crid}
\ee
They define a complex function in the variable $z=\s+\rmi t$,
\be\label{f}
f(z)\equiv u+\rmi v = \tan \frac{z}{2}\,.
\ee
Since $\p_\s v$ is odd in $\s$, it is possible to write the OSFT solution in Eq.~\Eq{regux} as ($t\ne 0$)
\be
\psi(r,t)= 3\int_{-\infty+\rmi t}^{+\infty+\rmi t} \rmd z\  K(z-\rmi t,r)\,\p_\s f(z)\,.\label{psizbos}
\ee
The BSFT solution Eq.~\Eq{trreg} in complex form reads
\be
\vp(r,t)=\frac{1}{2}-\frac{\rmi}{2}\int_{-\infty+\rmi t}^{+\infty+\rmi t}\rmd z\, K(z-\rmi t,r)\,f(z)\,,
\label{vpzbos}
\ee
where we exploited the $\s$ parity of $u$. From the Cauchy--Riemann identity $\p_\s f=-\rmi\p_t f$ one gets Eq.~\Eq{bcc}.

It should be remarked that this is a relations between exact solutions (BSFT) and approximate solutions of an approximate equation (OSFT). Consequently, at first sight it could sound incidental. However, there are several arguments supporting the existence of a correspondence between the two theories.
An independent argument underlying a relation between bosonic BSFT and OSFT was advocated in \cite{CST,ell07}. In \cite{ell07}, Ellwood showed that OSFT and BSFT solutions are related by a finite gauge transformation. In our case such a transformation should be parametrized by $r$. At present it is not clear if there is a precise quantitative relation between Ellwood's gauge transformation and the BSFT/OSFT correspondence outlined here. 

In this work, we have regarded the $(1+1)$-dimensional nature of the tachyon field as just a mathematical trick to localize the effective equation of motion of OSFT. A varying parameter $r$ is the key factor to link the solutions of cubic SFT with those of boundary SFT. According to this correspondence, the parameter $r$ allows one to interpolate between the (unit disk) bulk where the worldsheet string field theory is conformal (OSFT) and the boundary where nonconformal interactions are turned on (BSFT) \cite{CN}. When $r>0$, both cubic and boundary solutions $\psi_-$ and $\vp_-$ present wild oscillations. The choice of $r$ is not facultative in OSFT, as is dictated by worldsheet conformal symmetry: this is the reason why the only possible cubic solution oscillates at higher and higher absolute values of the potential. Also parity-definite solutions, if they are granted validity, are pathological for ${\rm sgn} (r)>0$, as they are bounded but nonanalytic at the origin. On the other hand, in BSFT $r$ is arbitrary and one can set it to 0 or to negative values; in such cases the BSFT solution, together with its derivative, is bounded and analytic. In parallel, even solutions with ${\rm sgn} (r)\leq0$ are bounded and smooth.

One encounters the same situation in the supersymmetric case, despite the lack of a simple relation like Eq.~\Eq{bcc}. 
Using 
\be
u(\s,t)=\frac{\cos\s}{\cosh t + \sin \s}
\,,\qquad
v(\s,t)=\frac{\sinh t}{\cosh t +\sin\s}\,,
\ee
the even susy OSFT solution is
\be
\psi(r,t)=\int_{-\infty+\rmi t}^{+\infty+\rmi t} \rmd z\  K(z-\rmi t,r)\,f\left(\frac{\pi}2-z\right)\,.\label{psizsus}
\ee
As for the bosonic tachyon, the integral of the solution is $r$-independent. 
However, the primitive 
is not the solution of supersymmetric BSFT Eq.~\Eq{superz4}. This can also be seen by looking at the integral form Eq.~\Eq{superz4}, which in complex notation reads
\be
\vp(r,t)=\frac{1}{2}-\frac{\rmi}{2}\int_{-\infty+\rmi t}^{+\infty+\rmi t}\rmd z\, K(z-\rmi t,r)\,f(\sqrt{2}z)\,.\label{vpzsus2}
\ee
This expression is similar to the OSFT solution but with the difference that the argument of the function convoluted with the kernel is shifted along the real axis and rescaled (reversing the time rescaling we made in Sec.~\ref{setup}, $z\to z/\sqrt{2}$ in Eq.~\Eq{psizsus}, the total relative rescaling factor is 2). In the series representation, the shift is responsible for the summation only over odd numbers in the exponents. Again, the argument is unchanged when considering the wildly oscillating solutions $\psi_-$ and $\vp_-$.


\section{Conclusions}\label{concl}

Let us summarize the main results of this paper.
\begin{enumerate}
\item[(i)] We found and discussed in detail approximated solutions to the fully nonlocal lowest-level equation of motion for the tachyon in supersymmetric open string field theory (in particular, in the 0-picture formulation). One solution is even and global, and has a spike at the origin which can be regularized or smoothened; this solution is eventually discarded. The other solution is related to that with increasing oscillations already studied in the literature and is valid up to some critical time. The description of these properties extends also, with minor modifications, to the bosonic case presented in \cite{FGN}. One can verify that the approximation $\rme^{r_*\B}\tphi^2\approx \tphi^2$ proposed to simplify the quartic potential is not valid for our solutions.
\item[(ii)] All these results stem from a method which can be of broader application in the general class of nonlocal theories. The study of nontrivial nonlocal cosmologies under the same procedure is in progress \cite{ctac2,CN}.
\item[(iii)] A family of exact solutions of boundary string field theory was found, both in the bosonic and supersymmetric case. When the ambiguity $r$ is positive, these solutions have wild oscillations, as in the OSFT case.
\item[(iv)] BSFT and OSFT solutions are formally related by a continuously varying parameter $r$ which takes fixed values in the physical case for each SFT. In the context of BSFT, this parameter can be naturally interpreted in two complementary ways: as the kernel of the Green function in the boundary action, and as a normalization or normal-ordering ambiguity in the boundary states corresponding to the open string partition function. The sign of $r$ determines the behaviour of these solutions in each theory; in particular, wild oscillations always occur in parity-undefinite solutions when $r>0$.
\end{enumerate}
This is a summary of wildly oscillating solutions ($r>0$):
\begin{itemize}
\item {\bf Bosonic OSFT:} Eq.~\Eq{psix} (series representation).
\item {\bf Bosonic BSFT:} Eq.~\Eq{q2box} (series representation).
\item {\bf Supersymmetric OSFT:} Eq.~\Eq{nloc3x} (series representation) or \Eq{irx} (integral representation). The solution related to the exact potential Eq.~\Eq{4pot} is the same but with $r\to 3r/2$ and a rescaled normalization.
\item {\bf Supersymmetric BSFT:} Eq.~\Eq{superz4} (series representation).
\end{itemize}
The case $r<0$ is allowed in BSFT and gives rise to bounded, smooth solutions: Eqs.~\Eq{trint} and \Eq{superz6} for bosonic and susy BSFT, respectively.

There are several issues which have not been considered here. Other tachyonic profiles may be chosen (e.g., \cite{JJKM}), as well as particular compactification schemes. Also, we have not given an explanation of the difference between the bosonic and susy correspondence. For the time being we notice that the supersymmetric cubic string field theory is less explored than its bosonic counterpart. The gauge trasformation of \cite{ell07} was derived explicitly only in the latter case; also, there are different proposals regarding the susy OSFT action. However, there seems to be no reason why the bosonic correspondence should not have a supersymmetric version; also, all susy OSFT candidates predict a local ($r=0$) lowest-level effective action for the tachyon with quadratic + quartic potential, which fixes the initial condition of the nonlocal problem. Other tachyon profiles in BSFT would unlikely account for the difference in the series coefficients. On the other hand the source of discrepancy might be traced in the different field dependence of the partition function (i.e., effective action) with respect to the tachyon profile $T$. In both the bosonic and susy case the BSFT renormalized tachyon field is $\vp=1-Z$ but the partition function is $Z\sim \rme^{-T}$ for the bosonic string \cite{KMM2,GS}, while $Z\sim \rme^{-T^2/4}$ for the susy string \cite{KMM1}. In this respect, it is not surprising to have found different relations for the two string theories. This issue will require further investigation; for the time being, the BSFT/OSFT relationship may be considered a technical device rather than a physical correspondence.


\begin{acknowledgments}
G.C. is supported by a Marie Curie Intra-European Fellowship under contract No.~MEIF-CT-2006-024523, and thanks I. Aref'eva and S. Panda for useful discussions. The authors thank V. Forini and G. Grignani for valuable comments and their initial involvement in this project. The work of G.N. is partly supported by INFN of Italy.
\end{acknowledgments}


\appendix


\section{Relations between different representations\newline of the bosonic BSFT solution}\label{bsrel}

In Sec.~\ref{mink}, we claimed that the series and integral representations [Eq.~\Eq{nloc3} and \Eq{ana}, respectively] of the OSFT supersymmetric solution ($r>0$) are equivalent, and both are related by analytic continuation to the integral representation Eq.~\Eq{nloc1} (rescaled) in the region $r<0$. Here we show this in the case of the analogous formul\ae\ of bosonic BSFT. Only the final result Eq.~\Eq{trreg} was presented in Sec.~\ref{bsftsol}, which we write again for convenience of the reader:
\be
\vp(r,t)=\frac{1}{2}+\frac{1}{2}\int_{-\infty}^\infty \rmd\s\, \frac{\rme^{-\frac{\s^2}{4r}}}{2\sqrt{\pi r}}\,\frac{\sinh y}{\cosh y + \cos\s}\,,\label{trregapp}
\ee
where $y=t+r+\ln T_0$. First, we find and discuss the series representation. When considering the $r\ne 0$ case, one has to apply the operator $\rme^{-r\p^2_t}$ to the solutions $\vp^{(\pm)} (0,y)$ of Eq.~(\ref{q1}). Since Eq.~(\ref{q1}) is an expansion of $\vp$ in terms of eigenfunctions of the operator $\p^2_t$, one would be tempted to replace $\rme^{-r\p_t^2}$ with its eigenvalue $\rme^{-rn^2}$ inside the sums, obtaining (for $r>0$)
\ba
\vp_+(r,t)=\sum_{n=0}^\infty (-1)^n \rme^{-rn^2} \rme^{-n y},\;\;\;\;
\;\;y>0\; ,\ \ \ r>0\,, \cr
\vp_-(r,t)=-\sum_{n=1}^\infty (-1)^n \rme^{-rn^2} \rme^{n y},\;\;\;\;
\;\; y<0\; ,\ \ \ r>0\,.\nn\\
\label{q2}
\ea
This choice corresponds to the strong limit and shows the discontinuity at the origin. In fact, 
\be 
\vp_+(r,0)-\vp_-(r,0)=\sum_{n=-\infty}^{\infty}(-1)^n \rme^{-rn^2}=\vartheta_4 (0,\rme^{-r})\ne 0\,,
\label{discont}
\ee
with $\vartheta_4 (u,q)=\sum_{n\in \mathbb{Z}} (-1)^n q^{n^2} \rme^{2\rmi n u}$ being the fourth Jacobi theta function. This discontinuity is troublesome because $\vp-1/2$ should be an antisymmetric function and as such it should vanish at the origin. 

Part of the troubles exhibited by Eq.~(\ref{q2}) at the origin are a consequence of the fact that the replacement $\rme^{-r\p_t^2}\to \rme^{-rn^2}$ implies an interchange of the order of two sums that indeed do not commute at $y=0$. Acting with the operator (\ref{lb}) on (\ref{q2}) leads to a double sum. Each of the sums over $n$ (for any fixed $\ell$) is divergent, and needs to be regularized. At $y=0$,
\ba
\left. \frac{}{} \vp_+\right|_{y=0} &=& 1+ \sum_{\ell=0}^\infty  \sum_{n=1}^\infty(-1)^n   \frac{(-r)^\ell}{\ell!}   n^{2 \ell}\cr
&=&1+ \sum_{\ell=0}^\infty  \frac{(-r)^\ell}{\ell!} (2^{2\ell + 1}-1) \zeta (-2\ell)\cr 
&=&\frac{1}{2}\ ,\cr
\left. \frac{}{} \vp_-\right|_{y=0} &= & - \sum_{\ell=0}^\infty  \sum_{n=1}^\infty(-1)^n   \frac{(-r)^\ell}{\ell!}   n^{2\ell}\cr
&=&-\sum_{\ell=0}^\infty  \frac{(-r)^\ell}{\ell!} (2^{2\ell + 1}-1) \zeta(-2\ell)\cr 
&=&\frac{1}{2}\,.
\label{tg2}
\ea
Consequently, the series representation provides the correct result $\vp^{(\pm)}(y=0)=1/2$ if the sums over $\ell$ and $n$ are not interchanged. This regularization at the origin characterizes the weak limit solution. In the integral representation it is encoded by a small regulator $\ve$ which smoothens the curve at the origin and is then set equal to 0 after integration. The discontinuity and the problem of the physical picture are removed either if $\ve\neq 0$ and the limit $\ve\to 0$ is not performed, or by taking $r\leq 0$.

A different way to understand why Eq.~(\ref{q2}) is problematic is the following. It would correspond to replace the operator $\rme^{-r\p_t^2}$ with $\rme^{-r\s^2}$ in the integrand of Eq.~(\ref{gammal}) and then closing with semicircles at infinity the contour $\Gamma_L$ to the right or to the left depending on the sign of $y$. However, this cannot be done, because when the factor $\rme^{-rs^2}$ is inserted in the integrand of (\ref{gammal}), the path $\Gamma_L$ cannot be closed by any curve at infinity, neither to the right nor to the left. In fact, if
$r<0$ the integral diverges at the points $\s=\pm \infty$, whereas if $r>0$ it diverges at $\s=\pm \rmi \infty$. Thus, the integral (\ref{gammal}) can never be computed as a sum of residues of the type (\ref{q2}). As already discussed, an appropriate regulator in the integral representation fixes the behaviour at the origin.

The demonstration that Eq.~\Eq{trregapp} (which has no regulator) and \Eq{q2} are equivalent goes as follows. We recast the integrand in the first equation as a Gaussian times
\be\label{trig}
\frac{\sinh y}{\cosh y+ \cos \s}=\Im \tan \frac{z}2=-1+\Im \left(\frac{2\rmi}{1+\rme^{\rmi z}}\right)\,,
\ee
where $z\equiv \s + \rmi y$. The first term gives a Gaussian integral which cancels the factor $1/2$ in Eq.~\Eq{trregapp}. If $y>0$, the last term in Eq.~\Eq{trig} is a convergent geometric series ($|\rme^{\rmi z}|=\rme^{-y}<1$):
\ba
\Im \left(\frac{2\rmi}{1+\rme^{\rmi z}}\right)&=& \Im\left[2 \rmi \sum_{n=0}^\infty (-1)^n \rme^{\rmi n z}\right]\cr
&=&2 \sum_{n=0}^\infty (-1)^n \rme^{-n y} \cos (n\s)\,.
\ea
Integration over $\s$ yields immediately $\vp_+$ in Eq.~\Eq{q2}. If $y<0$, one writes $\tan (z/2)$ in terms of $\rme^{-\rmi z}$ and repeats the same procedure to get $\vp_-$.

Now we would like to analytically continue Eq.~\Eq{trregapp} or \Eq{q2} to the region $r<0$. For instance, to show that Eq.~\Eq{trint} is the analytic continuation of the solution with $r>0$, we take one of the two branches of Eq.~(\ref{q2}), say $\vp_+$, and use the relation (valid for $n> 0$)
\be
\rme^{-r n^2}=\mp \frac{1}{2\pi \rmi}\int_{-\infty}^{+\infty}\rmd \s \frac{\rme^{\mp \rmi r
\s n}}{\s \pm \rmi n}\,,~~~~~~~r>0\,.
\ee
By the residue theorem, we can then rewrite $\vp_+$ as
\ba
\vp_+ &=& 1\mp \frac{1}{2\pi \rmi}\sum_{n=1}^\infty (-1)^n \rme^{-n y}\int_{-\infty}^{+\infty}\rmd \s \frac{\rme^{\mp \rmi r\s n}}{\s \pm \rmi n}\nonumber\\
&=& 1\pm\frac{1}{(2 \pi \rmi)^2}\int_{-\infty}^{+\infty}\rmd\s\int_{\Gamma}\rmd s \frac{\pi}{\sin \pi s}\frac{\rme^{(\mp \rmi r\s-y)s}}{\s\pm \rmi s}\,,\nn\\ \label{mel}
\ea
where the contour $\Gamma$ is made of an upwards-oriented curve parallel to the imaginary $s$ axis, lying between the poles in $s=0$ and $s=1$, and closed at infinity on the right. Closing the path $\Gamma$ on the left, one finds 
\ba
\vp &=&1\pm\frac{1}{2\pi \rmi}\int_{-\infty}^{+\infty}\rmd \s\sum_{n=0}^\infty(-1)^n \frac{\rme^{(\pm \rmi  r\s+y)n}}{\s\mp \rmi n \pm \rmi \eta}\nn\\
&& \mp \frac{1}{2\pi \rmi}\int_{-\infty}^{+\infty}\rmd
\s\frac{\pi \rme^{r\s^2}\rme^{\mp \rmi y\s}}{\sinh
\pi\s}\,,\label{boscontin}
\ea
where $\eta>0$ is an arbitrary small constant which regularizes the integral for $n=0$. When $r<0$, the integrals in the second term of Eq.~\Eq{boscontin} vanish for $n\geq 1$, while the third term converges. Then 
\be
\vp=\frac{1}{2}+\frac{1}{2}\int_{-\infty}^{+\infty} \rmd\s\,\rme^{r \sigma^2}\,\frac{\sin(\s y)}{\sinh(\pi\s)}\,,\qquad r<0\,.
\ee
This completes the proof that the BSFT solution with $r>0$ [Eq.~\Eq{trregapp} or \Eq{q2}, which are equivalent] and the solution with $r<0$ [Eq.~\Eq{trint}] are one the analytic continuation of the other.


\section{Bosonic boundary states}\label{bounda}

In this section we shall construct the boundary states corresponding to the open string partition function of Sec.~\ref{bsftsol}. Let us first briefly review standard boundary states for the rolling tachyon with $r=0$. One begins 
with the Wick rotated profile $\rme^{\rmi X^0}$ which defines a conformal field theory with a marginal boundary interaction. One can then consider the theory as compactified on a circle of self-dual critical radius $R_{X^0}=1$. It is well-known that at this radius the normal-ordered operator $:\rme^{2\rmi X_L(z)}:$ in the left-moving sector forms a level-1 Kac--Moody $su(2)$ algebra together with $:\rme^{-2\rmi X_L(z)}:$ and $\rmi\p_z X_L(z)$. From now on, $X\equiv X^0$ and $X(z,\bar z)$ has to be considered a closed-string variable, $z=\rme^{\rmi(\tau+\s)}$, where $\tau$ and $\s$ are Euclidean worldsheet coordinates. In general $X(z,\bar z)$ is 
\ba
X(z,\bar z)&=&X(\tau,\s)\cr
&=&x+p\s+\frac{p_L-p_R}{2}\tau\cr
&&+\frac{\rmi}{\sqrt{2}}\sum_{m\ne 0}\frac{\rme^{-\rmi m \s}}{m}\left(\alpha_m \rme^{-\rmi m\tau}
+\tilde\alpha_m \rme^{\rmi m\tau}\right)\,,\nn\\
\ea
where $x$ is the center-of-mass coordinate of the string. At the self-dual radius $R_X=1$, $p_L=p_R$. It is useful to define the positive and negative parts of $X$ at $\s=0$ where the boundary state is inserted:
\ba
&&X_>(\tau)=\frac{\rmi}{\sqrt{2}}\sum_{m> 0}\frac{1}{m}\left(\alpha_m \rme^{-\rmi m\tau}
+\tilde\alpha_m \rme^{\rmi m\tau}\right)\,,\cr
&&
X_<(\tau)=\frac{\rmi}{\sqrt{2}}\sum_{m<0}\frac{1}{m}\left(\alpha_m \rme^{-\rmi m\tau}
+\tilde\alpha_m \rme^{\rmi m\tau}\right)\,.
\label{posneg}
\ea
Their commutator is
\be
[X_>(\tau_1),X_<(\tau_2)]=-\frac{1}{2}\ln\left[4\sin^2\left(\frac{\tau_1-\tau_2}{2}\right)\right]\,.
\ee

One also introduces the normal-ordered currents
\ba
&&J^1 = \half (J^+ + J^-) = \half (:\rme^{2\rmi X}: + :\rme^{-2\rmi X}:)\ ,\cr
&&
J^2 = \frac{1}{2\rmi} (J^+ - J^-) = \frac{1}{2\rmi} (:\rme^{2\rmi X}: - :\rme^{-2\rmi X}:)\ ,
\cr
&&J^3 = \rmi\p_z X\,,
\label{currents}
\ea
whose Laurent modes
\be  
J^i_n=\oint \frac{\rmd z}{2\pi \rmi} z^n J^i(z)
\ee
satisfy the level-1 Kac--Moody algebra 
\be
[J^i_m,J^j_n]=\frac{\delta^{ij}}{2} m \delta_{m+n,0}+\rmi \epsilon^{ijk}J^k_{m+n}\,,
\ee
which in turn implies
\ba
&&\left[J^3_n, J^3_m \right] = \frac{n}{2} \delta_{n+m,\,0}, \quad
\left[J^+_n, J^-_m \right] = 2J^3_{n+m} + n\delta_{n+m,\,0},\cr
&&\left[J^3_n, J^{\pm}_m \right]= \pm J^{\pm}_{n+m}. \nn
\ea
The tachyon profile corresponds to an $su(2)$ generator given by the zero mode of the $J^+$ current:
\be
J^+_0=\int_0^{2\pi} \frac{\rmd\tau}{2\pi} :\rme^{2\rmi X_L(\tau,\s=0)}: \equiv\oint \frac{\rmd z}{2\pi\rmi} :\rme^{2\rmi X_L(z)}:
\label{jp0}
\ee
(no Jacobian is needed when changing variable from $\tau$ to $z$ because we are integrating a weight-1 field).
The $su(2)$ algebra corresponding to the zero modes of these currents plays an important role when one constructs the boundary states for the conformal field theory with the periodic boundary interaction. The Neumann boundary state 
for the unperturbed $D$-brane can be represented in terms of the Ishibashi state of $su(2)$ as~\cite{ish89,CKLM}
\be
|N \rangle = \sum_j \sum_{m\ge 0} |j, m, -m\rangle\rangle\,,
\ee
where $|j,m,-m\rangle\rangle$ is the Virasoro--Ishibashi state for the primary $|j,m,m\rangle$. At the self-dual radius where the left and right momenta $p_L,\ p_R$ are equal, the boundary state $|B \rangle$ generated by the periodic boundary interaction can be obtained by acting with the $SU(2)$ group element $\rme^{\rmi T_0 J^+_0}$ on the Neumann boundary state  
\be \label{bound}
|B \rangle = \exp\left[\rmi T_0\oint \frac{\rmd z}{2\pi \rmi} :\rme^{\rmi X(z)}:
\right]|N\rangle = \exp\left(\rmi T_0 J^+_0\right)|N\rangle\,,
\ee
where the last step follows from the Neumann condition $X_L|N\rangle=X_R|N\rangle$. When the boundary interaction $T_0$ is turned off, the boundary state reduces to the Neumann state. It is known from earlier works \cite{CKLM,CK} that such a boundary state can be written in terms of the spin-$j$ representation matrix
of the rotation in the $J_z$ eigenbasis:
\be
|B\rangle=\sum_{j=0,1/2,\dots}\sum_{m=0}^j {\cal D}^j_{m,-m}|j,m,m\rangle\rangle\,,
\label{B}
\ee
where ${\cal D}^j_{m,-m}$ is the rotation matrix element
\ba
{\cal D}^j_{m,-m}&=&\langle j,m|\rme^{\rmi T_0 J^+_0}|j,-m\rangle=\langle j,m|\frac{(\rmi T_0 J^+_0)^{2m}}{(2m)!}|j,-m\rangle\cr
&=&\left(\begin{array}{c}
  j+m \\
  2m
\end{array}\right)(\rmi T_0)^{2m}.
\ea
This matrix element requires $m$ to be non-negative.

To obtain an even more explicit form for the boundary state $|B\rangle$, one can use the observation by Sen~\cite{se021} that the Virasoro--Ishibashi state $|j;m,m\rangle\rangle$ in Eq.~(\ref{B}) is built over the primary state $|j;m,m\rangle$ which, in this $c=1$ conformal field theory, has momentum $2m$ and therefore is created by a vertex of the form $:\rme^{2\rmi mX(\tau=0)}:_N$. Here the operator $X(\tau)$ is defined where the boundary state is inserted, at $\s=0$. Since it should describe a Neumann boundary state, it has to be constructed with the Neumann normal ordering $:\ :_N$, defined as~\cite{gab04}
\be
:\rme^{\rmi\omega X(\tau)}:_N\equiv \rme^{2 \rmi \omega X_<(\tau)} \rme^{\rmi\omega x}\rme^{\rmi\omega [X_<(\tau)-X_>(\tau)]}.
\ee
The exponent on the right annihilates the Neumann boundary state $|N\rangle$, $[X_<(\tau)-X_>(\tau)]|N\rangle=0$, and
for the one- and two-point functions one finds
\ba
&&\langle0|:\rme^{\rmi\omega X(\tau)}:_N|N\rangle=\delta(\omega)\,,\cr
&&\langle0|:\rme^{\rmi\omega X(\tau_1)}:_N\, :\rme^{\rmi\omega X(\tau_2)}:_N|N\rangle=\delta(\omega_1+\omega_2)\cr
&&\qquad\qquad\qquad\qquad\times\left[4 \sin^2\left(\frac{\tau_1-\tau_2}{2}\right)\right]^{-{\omega_1^2}}.\label{neum}
\ea
However, the primary state $|j;m,m\rangle$ has conformal weight $(j^2,j^2)$, and it can be obtained from $:\rme^{2\rmi mX(0)}:_N|0\rangle_c$  by acting on it with an operator ${\cal O}_{j,m}$ which is a combination of oscillators of total level $j^2-m^2$. Here, $|0\rangle_c$ is the $SL(2,\mathbb{C})$ invariant Fock vacuum for the closed string. This primary state can be expressed in the form
\be
|j;m,m\rangle=\rme^{\rmi\theta(j,m)}{\cal O}_{j,m} :\rme^{2\rmi mX(0)}:_N|0\rangle_c\,,
\label{senprim}
\ee
where $\theta(j,m)$ is a suitable phase.

The general expression for $|B\rangle$ is quite complicated except in the scalar sector, which does not involve any $X$ oscillator. Writing the boundary state in an expansion in the bosonic oscillator basis and performing a Wick rotation, one has
\be
|B\rangle=f(t)|0\rangle_c+g(t)\alpha_{-1}\tilde\alpha_{-1}|0\rangle_c+\dots\,,
\ee
where
\ba
f(t)|0\rangle_c&=&\sum_{j=0,1/2,\dots}(\rmi T_0)^{2j}|j;j,j\rangle\cr
&=&\sum_{j=0,1/2,\dots}(-T_0 \rme^{t})^{2j}|0\rangle_c\cr
&=&\frac{1}{1+T_0\rme^{t}}|0\rangle_c\,.
\label{fx0}
\ea
In the second of these equalities we have used Eq.~(\ref{senprim}) and the phase convention of Ref.~\cite{se021}, $\rme^{\rmi\theta(j,j)}=\rmi^{2j}$. The last equality provides the partition function found in~\cite{lar02} which corresponds to the case $r=0$, Eq.~(\ref{larsen}).

We shall now introduce the ambiguity $r$ in the boundary state formalism. The variable $r$ is related to the normal-ordering ambiguity which is present both in the Neumann normal-ordered vertex entering the primary state (\ref{senprim}) and in the definition of the currents $J^{\pm}$ of Eq.~(\ref{currents}).

The normal ordering in the currents of Eq.~(\ref{currents}) may contain a constant related to the prescription used to regularize the vertex. We shall now look for a prescription that provides a boundary state 
consistent with the open-string partition function computed in the previous section. This would amount in introducing a generic parameter in the normalization of the vertex.
Consider the vertex operator $:\rme^{\rmi\omega X(z,\bar z)}:$ at $\s=0$; its holomorphic part with $\omega=2$ provides the generator $:\rme^{2\rmi X_L(\tau,0)}:$ appearing in Eq.~(\ref{jp0}), which then enters the definition of the boundary state $|B\rangle$, Eq.~(\ref{bound}). A regularized version of this current with a normalization containing the conformal weight of the operator is given by
\be\label{vert}
V_\omega(\tau)=\rme^{\omega^2r/2}\rme^{\rmi\omega X_<(\tau)}\rme^{\rmi\omega x}\rme^{\rmi\omega X_>(\tau)}\,,
\ee
where $r$ is an arbitrary constant. The one-point function on the closed-string vacuum for this vertex is a $\delta$ function
\be
\langle0|V_\omega(\tau)|0\rangle_c=\delta(\omega)\,,
\ee
and the two-point function reads
\ba
\langle0|V_{\omega_1}(\tau_1)V_{\omega_2}(\tau_2)|0\rangle_c&=&\delta(\omega_1+\omega_2)\rme^{r{\omega_1^2}}\cr
&&\times\left[4 \sin^2\left(\frac{\tau_1-\tau_2}{2}\right)\right]^{-\frac{\omega_1^2}{2}}.\nn\\
\label{2point}
\ea
This correlation function reproduces the propagator structure of Eq.~\Eq{prop0}. The current $:\rme^{2 \rmi X_L(z)}:$ should be normalized with half of the factor in Eq.~\Eq{vert}, becoming $\rme^{r}:\rme^{2 \rmi X_L(z)}:$. As a consequence, to preserve the Kac--Moody algebra unchanged the generator $:\rme^{-2 \rmi X_L(z)}:$ should become $\rme^{-r} :\rme^{-2 \rmi X_L(z)}:\,$.

One can define the vertex \Eq{neum} with a suitable normalization which provides the boundary correlators 
for a Neumann open string coordinate with the propagator prescription \Eq{prop0}. This should be given by
\be
\rno\rme^{\rmi\omega X(\tau)}\rno\equiv \rme^{r\omega^2}:\rme^{\rmi\omega X(\tau)}:_N\,.
\ee
In terms of this operator, the correlation functions read
\ba
&&\langle0|\rno \rme^{\rmi\omega X(\tau)}\rno|N\rangle=\delta(\omega)\,,\cr
&&\langle0|\rno\rme^{\rmi\omega X(\tau_1)}\rno\ \rno\rme^{\rmi\omega X(\tau_2)}
\rno|N\rangle=\delta(\omega_1+\omega_2)\cr
&&\qquad\qquad\qquad\times\left[\rme^{-2r}\sin^2\left(\frac{\tau_1-\tau_2}{2}\right)\right]^{-{\omega_1^2}},
\ea
and are consistent with the open string correlation functions computed in Sec.~\ref{bsftsol}. Therefore, the normalization for the bulk vertex giving the current generating the Kac--Moody algebra and that for the boundary vertex operator entering in Eq.~(\ref{senprim}) are different.

Then, the boundary state becomes
\ba
|B \rangle &=& \exp\left(\rmi T_0\rme^{r} J^+_0\right)|N\rangle\cr
&=&\sum_{j=0,1/2,\dots}\sum_{m=0}^j {\cal D}^j_{m,-m}|j,m,m\rangle\rangle\,,\label{boundq}
\ea
where ${\cal D}^j_{m,-m}$ now is
\be
{\cal D}^j_{m,-m}=\langle j,m|\rme^{\rmi T_0\rme^{r} J^+_0}|j,-m\rangle=\left(\begin{array}{c}
  j+m \\
  2m
\end{array}\right)\left(\rmi T_0\rme^{r}\right)^{2m}\,.
\ee
The primary state (\ref{senprim}) becomes
\be
|j;m,m\rangle=\rme^{\rmi\theta(j,m)}{\cal O}_{j,m} \rme^{-r(2m)^2}\rno\rme^{2\rmi mX(0)}\rno|0\rangle_c\,.
\ee
Expanding the boundary state in the bosonic oscillator basis and performing a Wick rotation, one finds $|B\rangle=f(t)|0\rangle_c+\dots$, where
\ba
f(t)|0\rangle_c&=&\sum_{j=0,1/2,\dots}\left(\rmi T_0\rme^{r}\right)^{2j}|j;j,j\rangle\cr
&=&\sum_{j=0,1/2,\dots}\left(-T_0\rme^{t+r}\right)^{2j} \rme^{-r(2j)^2}|0\rangle_c\cr
&=&\sum_{n=0}^\infty (-1)^n \rme^{-rn(n-1)} T_0^n \rme^{n t}|0\rangle_c\,.\label{laste}
\ea
This is precisely the partition function $1-\vp_-$ for the general case $r\ge 0$ found in Appendix \ref{bsrel}. 
The radius of convergence of this series is infinite (respectively, zero) for $r>0$ ($r<0$). The sign of $r$ is determined by the choice of writing the partition function as a perturbative series, but we have seen how to find representations of $Z$ valid also for $r<0$.

For $r=0$, Eq.~\Eq{laste} reproduces Eq.~(\ref{fx0}), otherwise one gets a solution with wild oscillations.


\end{document}